\documentclass[pra,preprint,superscriptaddress,floatfix,showpacs]{revtex4-1}

\usepackage{amsmath,amsfonts,amssymb}
\usepackage{graphicx}

\def\beq{\begin{equation}}
\def\eeq{\end{equation}}
\def\beqn{\begin{eqnarray}}
\def\eeqn{\end{eqnarray}}

\def\bi {\mbox{\boldmath $i$}}
\def\bo {\mbox{\boldmath $1$}}

\def\x {{\bf x}}
\def\y {{\bf y}}
\def\z {{\bf z}}
\def\p {{\bf p}}
\def\Q {{\bf Q}}
\def\P {{\bf P}}

\def\X {{\bf X}}
\def\Y {{\bf Y}}
\def\Z {{\bf Z}}

\begin{document}

\title{Properties of a trapped multiple-species bosonic mixture at the infinite-particle-number limit: A solvable model}

\author{O. E. Alon}\email{ofir@research.haifa.ac.il}
\affiliation{Department of Physics, University of Haifa, 3498838 Haifa, Israel}
\affiliation{Haifa Research Center for Theoretical Physics and Astrophysics, University of Haifa, 3498838 Haifa, Israel}
\author{L. S. Cederbaum}
\affiliation{Theoretical Chemistry, Physical Chemistry Institute, Heidelberg University, D-69120 Heidelberg, Germany}

\begin{abstract}
Trapped mixtures of Bose-Einstein condensates have attracted much attention.
The properties of such mixtures are rather difficult to compute and analyze.
This statement applies the more, the more species are involved.
We investigate a trapped mixture of Bose-Einstein condensates consisting of a multiple number of P species.
To be able to do so, an exactly-solvable many-body model is called into play.
This is the $P$-species harmonic-interaction model.
After presenting the Hamiltonian, the ground-state energy and wavefunction are explicitly calculated.
The solution is facilitated by utilizing a double set of Jacoby coordinates,
the first is a set of relative-motion Jacoby coordinates for each of the species and the second is a set of Jacoby coordinates built
from the center-of-mass coordinates of the $P$ species.
All properties of the mixture's ground state can in principle be obtained from the many-particle wavefunction.
A scheme to integrate the all-particle density matrix is derived and implemented,
leading to closed-form expressions for the reduced one-particle density matrices.
Of particular interest is the infinite-particle-number limit,
which is obtained when the numbers of bosons are taken to infinity while keeping the interaction parameters fixed.
We first prove that at the infinite-particle-number limit {\it all} the species are $100\%$ condensed.
The mean-field solution of the $P$-species mixture is also obtained analytically,
and is used to show that the energy per particle and densities per particle computed
at the many-body level of theory boil down to their mean-field counterparts.
Despite these, correlations in the mixture exist at the infinite-particle-number limit.
The availability of the solvable model allows one to enter an uncharted ground and explore how correlations of a
bosonic mixture at the infinite-particle-number limit depend on the interactions and, in particular, on the number of species $P$ in the mixture.
To this end, we obtain closed-form expressions for the correlation energy, namely, the difference between the mean-field and many-body energies,
and the depletion of the species, i.e., the number of particles residing outside the condensed modes, at the infinite-particle-number limit.
The depletion and the correlation energy per species are shown to critically depend on the number of species.
Of separate interest is the entanglement between one species of bosons and the other $P-1$ species.
This quantity is governed by the coupling of the center-of-mass coordinates of the species and is obtained by the respective Schmidt decomposition
of the $P$-species wavefunction.
Interestingly, there is an optimal number of species, here $P=3$, where the entanglement is maximal.
Importantly, the manifestation of this interspecies entanglement in an observable is possible.
It is the position-momentum uncertainty product of one species in the presence of the other $P-1$ species which is derived and demonstrated
to correlate with the interspecies entanglement.
All in all, we show and explain how correlations at the infinite-particle-number limit of a trapped multiple-species bosonic mixture
depend on the interactions, and how they evolve with the number of species.
Generalizations and implications are briefly discussed.
\end{abstract}

\maketitle 

\section{Introduction}\label{INTRO_SEC}

There has been a long interest in the properties of quantum systems made of different types of identical particles,
see, e.g., \cite{book1,book2,book3,book4}.
In the context of ultracold quantum gases \cite{RMP1,RMP2,RMP3},
mixtures of bosons
with a multiple
number of species
are attracting nowadays increased attention, see, e.g., Refs.~\cite{mmix1,mmix2,mmix3,mmix4,mmix5,mmix6,mmix7,mmix8,mmix9,mmix10}.
It goes without saying that the increased complexity of a multiple-species quantum system enriches its properties.
These are the kind of many-particle systems and research questions
we are interested at in the present work.

The ability to describe the dynamics and even the ground state of multiple-species
bosonic mixtures is obviously more challenging than for single-species bosons.
Whereas mean-field theory retains its structure as a product state when increasing the number of particles and species in the mixture,
a growing amount of approximations are generally a must at the many-body level,
as the complexity of the many-body wavefunction increases with the number of particles and species.
In this respect, the approach being pursued here is complementary.
We present and discuss a solvable many-particle model for a trapped multiple-species mixture of Bose-Einstein condensate,
where all particles interact by harmonic forces and trapped in an harmonic potential.
The model is solved throughout at the many-body level of theory,
and properties are computed from the many-particle wavefunction
and its reduced one-particle density matrices.
Also, for the sake of analysis, 
the model is solved analytically at the mean-field level of theory, see below.
We mention that 
harmonic-interaction models for systems of identical particles -- bosons, fermions, and mixtures -- have drawn continuous attention,
see, e.g., Refs.~\cite{HM1,HM2,Robinson_1977,HM4,HM5,HIM_Cohen,HIM_MIX_IJQC,HM8,HM9,HM10,
HM11,HM12,HM13,Schilling_2013,EPJD,HM16,HM17,HM18,HIM_MIX_RDM,HIM_BEC_CAVITY}.

An intriguing facet in the theory of single-species trapped Bose-Einstein condensates is the so-called infinite-particle-number limit.
In this limit, the product of the number of bosons and the strength of interaction between them is held fixed,
while the number of bosons is increased to infinity.
Then, it can be proved mathematically rigorously that
certain many-body properties precisely boil down to their mean-field counterparts.
These chiefly include the ground-state energy per particle, density per particle, and 100\% condensate fraction,
i.e., the leading eigenvalue of the reduced one-particle density matrix per particle \cite{INF1,INF2}, also see \cite{Castin,INF_LENZ_2017}.
On the other hand, there are many-body properties of single-species trapped Bose-Einstein condensates
that do not coincide at the limit of an infinite number of particles with their mean-field counterparts.
These include variances per particle of observables, like the many-particle position, momentum and angular-momentum operators,
that depend on the number of depleted particles, which is non-zero, rather than on the depleted fraction, which, as said, vanishes
at the infinite-particle-number limit \cite{INF3,INF4}.
Ultimately, the many-body and mean-field wavefunctions are different from each other,
their overlap can be much smaller than unity \cite{INF_LENZ_2017,INF_OVERLAP_2016},
and, consequently, properties derived from them can differ,
even at the limit of an infinite number of particles.

For mixtures,
there are relatively
more recent studies
on their properties at the infinite-particle-number limit.
To start with,
also for mixtures it has been shown under wide conditions that the 
ground-state energy per particle and density per particle of each species
boil down to the predictions of mean-field theory for mixtures,
and that each of the species is 100\% condensate \cite{HIM_MIX_RDM,BB1,BB2,BB3,BB4}.
On the other hand, just as described above for single-species bosons,
variances per particle of observables in a mixture computed at the many-body and mean-field
levels of theory can deviate substantially from each other \cite{HIM_MIX_CP}.
Interestingly and beyond that,
one may expect that there are properties for mixtures at the infinite-particle-number limit 
that cannot or do not occur for single-species bosons in this limit.
For example,
the entanglement between two species can remain finite at the infinite-particle-number limit,
despite 100\% condensation of the species \cite{Atoms_2021}.
Furthermore,
in a periodically-driven mixture,
the quasienergy per particle computed at the many-body level of theory need not coincide 
with the respective mean-field quasienergy,
despite 100\% condensation of the driven species \cite{Floquet_HIM}.
This is unlike the above-mentioned property of the ground-state energy of mixtures.
For sure,
the number of species itself in the mixture is expected to be an interesting `parameter' that,
as we shall demonstrate and explore in the present work,
can govern correlations at the infinite-particle-number limit of the mixture.

The contents of the paper is as follows.
In Sec.~\ref{BBB_3_SEC}, we treat the triple-species mixture
and start building the tools needed to treat multiple-species mixtures with a general number of $P$ species.
In Sec.~\ref{BBB_P_SEC}, we generalize the derivation
to multiple-species mixtures.
In both sections, the derivations are made for a mixture with a
finite number of particles,
then the infinite-particle-limit is taken
and closed-form expressions of properties in this limit are obtained.
In Sec.~\ref{SEC_INVESTIGATION},
a detailed investigation of correlations in a $P$-species mixture
at the infinite-particle-number limit
is performed,
and the dependence of these correlations
on interactions and, in particular, on the number of species
in the mixture is elucidated.
Section \ref{SEC_SUM_OUT} is devoted to a summary and outlook.
Finally,
the appendices present complimentary materials.
Appendix \ref{SEC_MIX2}
collects and augments literature results obtained using
the harmonic-interaction model for two-species mixtures,
as far as they are needed for comparisons with and analysis of the main text.
Finally, appendix \ref{SEC_FOLD} proves by induction the folding of the
center-of-mass coordinates in the $P$-species mixture.

\section{The smallest multiple-species mixture: Three trapped Bose-Einstein condensates}\label{BBB_3_SEC}

Let the many-particle Hamiltonian be
\beqn\label{HAM_MIX_3}
& & \hat H(\x_1,\ldots,\x_{N},\y_1,\ldots,\y_{N},\z_1,\ldots,\z_{N}) = 
\sum_{j=1}^{N} \left( -\frac{1}{2} \frac{\partial^2}{\partial \x_j^2} + \frac{1}{2}\omega^2 \x_j^2 \right) +
\nonumber \\
& & + 
\sum_{j=1}^{N} \left( -\frac{1}{2} \frac{\partial^2}{\partial \y_j^2} + \frac{1}{2}\omega^2 \y_j^2 \right) + 
\sum_{j=1}^{N} \left( -\frac{1}{2} \frac{\partial^2}{\partial \z_j^2} + \frac{1}{2}\omega^2 \z_j^2 \right) +
\nonumber \\
& & + \lambda_1 \sum_{1 \le j < k}^{N} (\x_j-\x_k)^2 + 
\lambda_1 \sum_{1 \le j < k}^{N} (\y_j-\y_k)^2 + 
\lambda_1 \sum_{1 \le j < k}^{N} (\z_j-\z_k)^2 + \nonumber \\
& &
+ \lambda_{12} \sum_{j=1}^{N} \sum_{k=1}^{N} (\x_j-\y_k)^2
+ \lambda_{12} \sum_{j=1}^{N} \sum_{k=1}^{N} (\x_j-\z_k)^2
+ \lambda_{12} \sum_{j=1}^{N} \sum_{k=1}^{N} (\y_j-\z_k)^2. \
\eeqn
It describes a balanced mixture of three Bose-Einstein condensates in a trap,
i.e., there are $N$ bosons in each species and the mass of each boson is the same,
taken to be one for convenience.
All bosons interact by harmonic forces and trapped in an harmonic potential.
Note, also, that in the balanced mixture the intraspecies interactions $\lambda_1$ are alike,
and the interspecies interactions $\lambda_{12}$ are the same.
It is useful to specify and express quantities using the interaction parameters
$\Lambda_1=\lambda_1(N-1)$ and $\Lambda_{12}=\lambda_{12}N$.
These will be used throughout when defining the limit of an infinite number of particles
and analyzing quantities in this limit.
Explicitly, $\Lambda_1$ and $\Lambda_{12}$ are held fixed while the number of bosons $N$ in each species is increased to infinity.
The Hamiltonian (\ref{HAM_MIX_3}) is a straightforward generalization
of the two-species mixture's Hamiltonian, see appendix \ref{SEC_MIX2}.

To diagonalize (\ref{HAM_MIX_3}), we start with 
the Jacoby coordinates for species $1$, $2$, and $3$,
\beqn\label{MIX_COOR_3}
& & \X_s = \frac{1}{\sqrt{s(s+1)}} \sum_{j=1}^{s} (\x_{s+1}-\x_j),
\ \ 1 \le s \le N-1, \quad \X_{N} = \frac{1}{\sqrt{N}} \sum_{j=1}^{N} \x_j, \nonumber \\
& & \Y_s = \frac{1}{\sqrt{s(s+1)}} \sum_{j=1}^{s} (\y_{s+1}-\y_j),
\ \ 1 \le s \le N-1, \quad \Y_{N} = \frac{1}{\sqrt{N}} \sum_{j=1}^{N} \y_j, \nonumber \\
& & \Z_s = \frac{1}{\sqrt{s(s+1)}} \sum_{j=1}^{s} (\z_{s+1}-\z_j),
\ \ 1 \le s \le N-1, \quad \Z_{N} = \frac{1}{\sqrt{N}} \sum_{j=1}^{N} \z_j, \
\eeqn
which consist of relative-motion and center-of-mass coordinates.
Then,
the Hamiltonian (\ref{HAM_MIX_3}) may be written as a sum of two Hamiltonians,
\beqn\label{HAM_MIX_3_XY}
& & \hat H = \hat H_{rels} + \hat H_{CMs}. \
\eeqn
The Hamiltonian of the relative motions is given by
\beqn\label{HAM_MIX_3_XYZ_rels}
& &
\hat H_{rels}(\X_1,\ldots,\X_{N-1},\Y_1,\ldots,\Y_{N-1},\Z_1,\ldots,\Z_{N-1}) =
\sum_{j=1}^{N-1} \left( -\frac{1}{2} \frac{\partial^2}{\partial \X_j^2} + \frac{1}{2} \Omega_1^2 \X_j^2 \right) +
\nonumber \\
& & + \sum_{j=1}^{N-1} \left( -\frac{1}{2} \frac{\partial^2}{\partial \Y_j^2} + \frac{1}{2} \Omega_1^2 \Y_j^2 \right)
+ \sum_{j=1}^{N-1} \left( -\frac{1}{2} \frac{\partial^2}{\partial \Z_j^2} + \frac{1}{2} \Omega_1^2 \Z_j^2 \right), \
\eeqn
where the intraspecies relative-motion frequencies are equal to
\beqn\label{rel_freq_BBB}
& & \Omega_1 = \sqrt{\omega^2+2\left[\left(1+\frac{1}{N-1}\right)\Lambda_1+2\Lambda_{12}\right]}. \
\eeqn
One may infer, despite the mixture being balanced, that
the frequency $\Omega_1$ originates from and 
depends on the interactions of, say, species $1$ with species $2$ and $3$,
but not on the interspecies interaction between the latter two.

The center-of-masses Hamiltonian can be written as follows,
\beqn\label{HAM_MIX_3_XYZ_CMs}
& &
\hat H_{CMs}(\X_{N},\Y_{N},\Z_{N}) =
- \frac{1}{2} \left(\frac{\partial^2}{\partial \X_{N}^2} + \frac{\partial^2}{\partial \Y_{N}^2}
+ \frac{\partial^2}{\partial \Z_{N}^2}\right)
+ \frac{1}{2}
\begin{pmatrix}
\X_{N} &
\Y_{N} &
\Z_{N} \cr
\end{pmatrix}
\underline{\underline{\bf O}}
\begin{pmatrix}
\X_{N} \cr
\Y_{N} \cr
\Z_{N} \cr
\end{pmatrix}, \nonumber \\
& &
\underline{\underline{\bf O}} = \left(\omega^2 + 6\Lambda_{12}\right) \underline{\underline{\bf I}}
-2\Lambda_{12}\underline{\underline{\bf 1}}, \
\eeqn
where $\underline{\underline{\bf I}}$ is the unit matrix and $\underline{\underline{\bf 1}}$ is the matrix with all ones.
Diagonalizing the frequencies' matrix $\underline{\underline{\bf O}}$
one finds the eigenvalues and eigenvectors emerging
from the center-of-mass degrees-of-freedom.
In the three-species mixture
there are two equal relative-coordinate frequencies in addition to the center-of-mass frequency,
\beqn\label{CM_3_freq}
& & \Omega_{123} = \sqrt{\omega^2 + 6\Lambda_{12}} \mathrm{\ \ (2 \ roots)}, \quad \omega, \
\eeqn
where
the corresponding eigenvectors are
\beqn\label{CM_3_vecs}
& &
\Q_1 = \frac{1}{\sqrt{2}}\left(-\X_N+\Y_N\right), \quad
\Q_2 = \frac{1}{\sqrt{6}}\left(-\X_N-\Y_N+2\Z_N\right), \quad
\Q_3 = \frac{1}{\sqrt{3}}\left(\X_N+\Y_N+\Z_N\right). \nonumber \\ \
\eeqn
Clearly, in the balanced mixture,
the components of the relative center-of-mass coordinates $\Q_1$ and $\Q_2$ do not depend on the interaction strengths,
like the respective relative coordinate of the two-species mixture [Eq.~(\ref{CM_2_vecs})].
Of course, the center-of-mass of the whole mixture does not depend on the interaction strengths.

The ground-state wavefunction then takes on the separable form
\beqn\label{WF_HIM_3}
& & \Psi(\X_1,\ldots,\Y_1,\ldots,\Z_1,\ldots,\Q_1,\Q_2,\Q_3) =
\left(\frac{\Omega_1}{\pi}\right)^{\frac{9(N-1)}{4}}
\left(\frac{\Omega_{123}}{\pi}\right)^{\frac{3}{2}}
\left(\frac{\omega}{\pi}\right)^{\frac{3}{4}} \times \nonumber \\
& & \times
e^{-\frac{1}{2} \left(\Omega_1 \sum_{k=1}^{N-1} \X_k^2
+ \Omega_1 \sum_{k=1}^{N-1} \Y_k^2
+ \Omega_1 \sum_{k=1}^{N-1} \Z_k^2\right)}
e^{-\frac{1}{2}\left(\Omega_{123} \Q_1^2 + \Omega_{123} \Q_2^2 + \omega \Q_3^2\right)}. \
\eeqn
Consequently,
the ground-state energy is
\beqn\label{E_HIM_3}
& & E = \frac{3}{2} \left[ 3(N-1) \Omega_1 + 2\Omega_{123} + \omega \right] = \nonumber \\
& &
= \frac{3}{2} \left[3(N-1) \sqrt{\omega^2+2\left[\left(1+\frac{1}{N-1}\right)\Lambda_1+2\Lambda_{12}\right]}
+ 2\sqrt{\omega^2 + 6\Lambda_{12}} + \omega \right]. \
\eeqn 
The energy is the sum of contributions of $3N$ oscillators,
all degrees-of-freedom but the center-of-mass are dressed by the interparticle interactions.
$E$ will be used to compute the correlation energy at the infinite-particle-number limit.
Of course, all frequencies must be positive for the three-species mixture to be bound, hence
\beqn\label{FREQ_3_BOUNDS}
 \Lambda_{12} > - \frac{\omega^2}{6}, \qquad
\Lambda_1 > - \left(1-\frac{1}{N}\right)\left(\frac{\omega^2}{2} + 2\Lambda_{12}\right).
\
\eeqn
The meaning of (\ref{FREQ_3_BOUNDS}) is that the
interspecies interactions are bound from below only by the trapping frequency,
whereas the intraspecies interactions are bound from below by the combination of the interspecies interactions and $\omega$.
Correspondingly,
the energy (\ref{E_HIM_3}) is bound from below but not from above.

When only intraspecies quantities are required, unlike the generic treatment of the two-species
mixture derived before \cite{HIM_MIX_RDM},
we do not need to transform to the laboratory-frame coordinates beforehand.
Instead, to integrate out the second and third species
we express the wavefunction using the 
Jacoby coordinates of each of the species,
\beqn\label{WF_HIM_3_JAC}
& & \Psi(\X_1,\ldots,\X_{N},\Y_1,\ldots,\Y_{N},\Z_1,\ldots,\Z_{N}) =
\left(\frac{\Omega_1}{\pi}\right)^{\frac{9(N-1)}{4}}
\left(\frac{\Omega_{123}}{\pi}\right)^{\frac{3}{2}}
\left(\frac{\omega}{\pi}\right)^{\frac{3}{4}} \times \nonumber \\
& & \times
e^{-\frac{1}{2} \Omega_1 \sum_{k=1}^{N-1} \X_k^2}
e^{-\frac{1}{2} \Omega_1 \sum_{k=1}^{N-1} \Y_k^2}
e^{-\frac{1}{2} \Omega_1 \sum_{k=1}^{N-1} \Z_k^2}
e^{-\frac{1}{2} a \X_{N}^2}
e^{-\frac{1}{2} a \Y_{N}^2}
e^{-\frac{1}{2} a \Z_{N}^2} \times \nonumber \\
& &
\times
e^{-b \X_{N}\Y_{N}}
e^{-b \X_{N}\Z_{N}}
e^{-b \Y_{N}\Z_{N}}, \
\eeqn
where the various coefficients of the
center-of-masses part of the wavefunction are
\beqn\label{WF_HIM_3_JACS_COEFF}
& &
a = \frac{1}{3}\left(2\Omega_{123}+\omega\right) = \Omega_{123}+b,  \qquad
b = \frac{1}{3}\left(\omega-\Omega_{123}\right). \
\eeqn
These coefficients are interrelated and satisfy
$a+2b = \omega$.
Furthermore,
$a$ and $b$
scale properly and remain finite at the limit of an infinite number of particles, see below.

We continue with the all-particle density matrix expressed using the species' Jacoby coordinates.
As said, taking a different path to the solution in \cite{HIM_MIX_RDM}
opens up the possibility of a tractable
integration scheme for the intraspecies reduced density matrices in the general multiple-species problem.
It also shorten the integration of the two-species mixture, see appendix \ref{SEC_MIX2}.
Thus, we have
\beqn\label{N1_N2_N3_DENS_JAC}
& & \Psi(\X_1,\ldots,\X_{N},\Y_1,\ldots,\Y_{N},\Z_1,\ldots,\Z_{N})
\Psi^\ast(\X'_1,\ldots,\X'_{N},\Y'_1,\ldots,\Y'_{N},\Z'_1,\ldots,\Z'_{N}) = \nonumber \\
& & =
\left(\frac{\Omega_1}{\pi}\right)^{\frac{9(N-1)}{2}}
\left(\frac{\Omega_{123}}{\pi}\right)^{3}
\left(\frac{\omega}{\pi}\right)^{\frac{3}{2}} \times \nonumber \\
& & \times
e^{-\frac{1}{2} \Omega_1 \sum_{k=1}^{N-1} \left(\X_k^2+{\X'}_k^2\right)}
e^{-\frac{1}{2} \Omega_1 \sum_{k=1}^{N-1} \left(\Y_k^2+{\Y'}_k^2\right)}
e^{-\frac{1}{2} \Omega_1 \sum_{k=1}^{N-1} \left(\Z_k^2+{\Z'}_k^2\right)}
e^{-\frac{1}{2} a \left(\X_{N}^2+{\X'}_{N}^2\right)}
\times \nonumber \\
& & 
\times
e^{-\frac{1}{2} a \left(\Y_{N}^2+{\Y'}_{N}^2\right)}
e^{-\frac{1}{2} a \left(\Z_{N}^2+{\Z'}_{N}^2\right)}
e^{-b \left(\X_{N}\Y_{N}+\X'_{N}\Y'_{N}\right)}
e^{-b \left(\X_{N}\Z_{N}+\X'_{N}\Z'_{N}\right)}
e^{-b \left(\Y_{N}\Z_{N}+\Y'_{N}\Z'_{N}\right)}, \
\eeqn
where the normalization here is taken to be one for simplicity.

To reduce the all-particle density matrix (\ref{N1_N2_N3_DENS_JAC}),
the integration scheme in the three-species mixture
begins with elimination of the relative-motion Jacoby coordinates of species $3$ and $2$,
proceeds over the center-of-mass of species $3$, $\Z'_{N}=\Z_{N}$, and that of species $2$, $\Y'_{N}=\Y_{N}$,
and gives
\beqn\label{INT_CM_BBB_3}
& &
\!\!\!\!\!\!\!\!
\int d\Y_{N} e^{-a \Y_{N}^2}
e^{-b \left(\X_{N}+\X'_{N}\right)\Y_{N}}
\int d\Z_{N} e^{-a \Z_{N}^2}
e^{-b\left[\left(\X_{N}+\X'_{N}\right)+2\Y_{N}\right]\Z_{N}}
= \nonumber \\
& &
\!\!\!\!\!\!\!\!
= \left(\frac{\pi}{a}\right)^{\frac{3}{2}} 
\left(\frac{\pi}{a-\frac{b^2}{a}}\right)^{\frac{3}{2}}
e^{+\frac{1}{4}\left[\frac{b^2}{a}+\frac{\left(b-\frac{b^2}{a}\right)^2}{a-\frac{b^2}{a}}\right]\left(\X_{N}+\X'_{N}\right)^2} =
\left(\frac{\pi^2}{a^2-b^2}\right)^{\frac{3}{2}}
e^{+\frac{1}{4}\left(\frac{2b^2}{a+b}\right)\left(\X_{N}+\X'_{N}\right)^2}. \
\eeqn
Comparison of the elimination of the center-of-masses of species $3$ and $2$ in the three-species mixture
[Eq.~(\ref{INT_CM_BBB_3})] to the respective elimination of species $2$ in the two-species mixture [Eq.~(\ref{INT_CM_BB_2})]
suggests a mechanism of how the center-of-mass of the remaining species $1$
gets dressed by the other species' center-of-masses.
We present the analysis and derive the explicit construction within the investigation
of the multiple-species mixture in the next section.

After eliminating species $3$ and $2$ from the all-particle density matrix (\ref{N1_N2_N3_DENS_JAC}),
the working expression for the reduced density matrix of the remaining species $1$ in the three-species mixture is
\beqn\label{F_1_0_0_BBB_3}
& & e^{-\frac{\alpha}{2} \sum_{j=1}^{N} \left(\x_j^2+{\x'}_j^2\right) - \beta \sum_{1\le j <k}^{N}
\left(\x_j \x_k + \x'_j \x'_k\right)} e^{-\frac{1}{4}C_{N,0,0}\left\{\sum_{j=1}^{N}\left(\x_j+\x'_j\right)\right\}^2}, \
\eeqn
with the three coefficients
\beqn\label{F_1_0_0_BBB_3_Coeff_1}
& & \alpha = \Omega_1 + \frac{1}{N}\left(a-\Omega_1\right) =
\Omega_1\left\{1+\frac{1}{N}\left[\frac{1}{3}\left(\frac{\omega}{\Omega_1}+\frac{2\Omega_{123}}{\Omega_1}\right)-1\right]\right\}, \nonumber \\
& &
\beta = \alpha-\Omega_1 = 
\Omega_1\frac{1}{N}\left[\frac{1}{3}\left(\frac{\omega}{\Omega_1}+\frac{2\Omega_{123}}{\Omega_1}\right)-1\right], \nonumber \\
& & C_{N,0,0} = - \frac{1}{N}\frac{2b^2}{a+b} =
- \frac{1}{N}\frac{2}{3} \frac{\left(\omega-\Omega_{123}\right)^2}{\left(2\omega+\Omega_{123}\right)}. \
\eeqn
Built by construction,
the three-species working expression (\ref{F_1_0_0_BBB_3}) has the same functional form like 
its two-species ancestor (\ref{F_1_0_BB_2}),
with analogous coefficients
that fold down exactly the effects of species $3$ and $2$ onto species $1$.

Further folding of $N-1$ coordinates,
the reduced one-particle density matrix and the one-particle density of species $1$ are given by
\beqn\label{rho_1_BBB_3}
& & \rho^{(1)}_1(\x,\x') = N \left(\frac{\alpha+C_{1,0,0}}{\pi}\right)^{\frac{3}{2}}
e^{-\frac{\alpha}{2}\left(\x^2+{\x'}^2\right)}e^{-\frac{1}{4}C_{1,0,0}\left(\x+\x'\right)^2}, \nonumber \\
& & \rho^{(1)}_1(\x) = N \left(\frac{\alpha+C_{1,0,0}}{\pi}\right)^{\frac{3}{2}}
e^{-\left(\alpha+C_{1,0,0}\right)\x^2}, \
\eeqn
where
\beqn\label{rho_1_BBB_3_coeff}
& &
\!\!\!\!\!\!\!\!\!\!
\alpha + C_{1,0,0} = \left(\alpha-\beta\right) \frac{\left(\alpha-\beta\right) + N\left(C_{N,0,0}+\beta\right)}
{\left(\alpha-\beta\right) + (N-1)\left(C_{N,0,0}+\beta\right)} = \Omega_1
\frac{1}{1+\frac{1}{N}\left[\frac{1}{3}\left(\frac{\Omega_1}{\omega}+\frac{2\Omega_1}{\Omega_{123}} \right) - 1 \right]}, \
\eeqn
with 
$C_{N,0,0} + \beta =
\frac{1}{N}\left(\frac{3\omega\Omega_{123}}{2\omega+\Omega_{123}} - \Omega_1\right) =
- \frac{3\omega\Omega_{123}}{2\omega+\Omega_{123}}
\frac{1}{N}\left[\frac{1}{3}\left(\frac{\Omega_1}{\omega}+\frac{2\Omega_1}{\Omega_{123}}\right)-1\right]$,
and where the auxiliary relation $C_{N,0,0} + \alpha = \Omega_1\left(1-\frac{1}{N}\right)+\frac{1}{N}\left[\frac{(a-b)(a+2b)}{a+b}\right] =
\Omega_1\left(1-\frac{1}{N}\right)+\frac{1}{N}\left(\frac{3\omega\Omega_{123}}{2\omega+\Omega_{123}}\right)$
is used.
We stress that expressions (\ref{rho_1_BBB_3}) with coefficient (\ref{rho_1_BBB_3_coeff}) are the
exact and explicit
reduced one-particle density matrix and density,
where all the interactions with and impact of species $2$ and $3$
are precisely folded onto
species $1$.

So far,
we have derived the exact wavefunction, energy, and the reduced one-particle density matrix and density
in the three-species mixture as a function of the numbers of particles and intraspecies
and interspecies interactions.
The explicit expressions are in principle ready for the investigation and analysis of finite systems,
which is a topic to be pursued elsewhere.
Since as an application we chose to construct the theory and investigate mixtures
at the infinite-particle-number limit,
we need additional ingredients to be obtained from the mean-field solution
of the Hamiltonian (\ref{HAM_MIX_3}).

The mean-field solution of the three-species mixture goes as follows.
The ansatz for the wavefunction is the separable product state
\beq\label{MIX_WAV_GP_3}
 \Phi^{GP}(\x_1,\ldots,\x_{N},\y_1,\ldots,\y_{N},\z_1,\ldots,\z_{N}) = 
\prod_{j=1}^{N} \phi^{GP}_1(\x_j) \prod_{k=1}^{N} \phi^{GP}_2(\y_k) \prod_{l=1}^{N} \phi^{GP}_3(\z_l). 
\eeq
No
assumptions are made on the shapes of the three orbitals,
aside their individual normalization to one.
Sandwiching the Hamiltonian (\ref{HAM_MIX_3}) with the ansatz (\ref{MIX_WAV_GP_3})
and
minimizing the resulting energy functional with respect to
the shapes of the orbitals $\phi^{GP}_1(\x)$, $\phi^{GP}_2(\y)$, and $\phi^{GP}_3(\z)$, 
the three coupled Gross-Pitaevskii equations of the mixture are derived,
{\small{
\beqn\label{MIX_EQ_GP_BBB_3}
& &
\!\!\!
\bigg\{-\frac{1}{2} \frac{\partial^2}{\partial \x^2} + \frac{1}{2} \omega^2 \x^2 
+ \int d\x' \big[\Lambda_1 |\phi_1(\x')|^2 + \Lambda_{12} \left(|\phi_2(\x')|^2 
+ |\phi_3(\x')|^2\right)\big] (\x-\x')^2\bigg\} \phi_1(\x) = \mu_1 \phi_1(\x),
\nonumber \\
& &
\!\!\!
\bigg\{-\frac{1}{2} \frac{\partial^2}{\partial \y^2} + \frac{1}{2} \omega^2 \y^2 
+ \int d\y' \big[\Lambda_1 |\phi_2(\y')|^2 + \Lambda_{12} \left(|\phi_1(\y')|^2
+ |\phi_3(\y')|^2\right)\big] (\y-\y')^2\bigg\} \phi_2(\y) = \mu_2 \phi_2(\y),
\nonumber \\
& &
\!\!\!
\bigg\{-\frac{1}{2} \frac{\partial^2}{\partial \z^2} + \frac{1}{2} \omega^2 \z^2 
+ \int d\z' \big[\Lambda_1 |\phi_3(\z')|^2 + \Lambda_{12} \left(|\phi_1(\z')|^2
+ |\phi_3(\z')|^2\right) \big] (\z-\z')^2\bigg\} \phi_3(\z) = \mu_3 \phi_3(\z). \nonumber \\ \
\eeqn
}}
The three coupled non-linear equations (\ref{MIX_EQ_GP_BBB_3}) are equivalent one to the others
in the balanced mixture.
Their solution is given by
\beqn\label{MIX_GP_OR_3}
& & \phi^{GP}_1(\x) = \left(\frac{\Omega_1^{GP}}{\pi}\right)^{\frac{3}{4}}
e^{-\frac{1}{2}\Omega_1^{GP}\x^2}, \quad
\Omega_1^{GP} = \sqrt{\omega^2 + 2(\Lambda_1+2\Lambda_{12})}, \nonumber \\
& & \phi^{GP}_2(\y) = \left(\frac{\Omega_1^{GP}}{\pi}\right)^{\frac{3}{4}}
e^{-\frac{1}{2}\Omega_1^{GP}\y^2},
\nonumber \\
& & \phi^{GP}_3(\z) = \left(\frac{\Omega_1^{GP}}{\pi}\right)^{\frac{3}{4}}
e^{-\frac{1}{2}\Omega_1^{GP}\z^2}, \
\eeqn
where $\mu_1=\mu_2=\mu_3=
\frac{3}{2}\left(\Omega_1^{GP}+\frac{\Lambda_1+2\Lambda_{12}}{\Omega_1^{GP}}\right)$
are the chemical potentials of the species.
Equivalently, $\mu_1$, $\mu_2$, and $\mu_3$ are the Lagrange multipliers ensuring the normalization of the orbitals 
$\phi^{GP}_1(\x)$, $\phi^{GP}_2(\y)$, and $\phi^{GP}_3(\z)$ to unity, respectively.
It is seen from (\ref{MIX_GP_OR_3}) 
that, in the balanced mixture, the orbitals of different species are the same.
Each species is dressed by the interactions with the other two,
but does not depend on the interaction between the other two species.
Furthermore,
the solution is symmetry preserving,
and no demixing in the present model occurs,
just as for the two-species mixture, see for further discussion appendix \ref{SEC_MIX2}.

The Gross-Pitaevskii energy per particle
of the three-species mixture reads
\beqn\label{MIX_GP_E_N_3}
& &
\varepsilon^{GP} = 
\frac{E^{GP}}{3N} =
\frac{3}{2}\Omega_1^{GP}=
\frac{3}{2}\sqrt{\omega^2 + 2\left(\Lambda_1+2\Lambda_{12}\right)},
\eeqn
marking that all bosons in the balanced three-species mixture contribute the
same amount to the energy at the mean-field level of theory.

The infinite-particle-number limit implies $100\%$ Bose-Einstein condensation and boiling down
to some of the mean-field quantities,
but this needs to be shown explicitly.
We start from $\lim_{N \to \infty \atop J=1,2,3} \Omega_1 = \Omega_1^{GP}$.
Consequently, one finds for the energy of the mixture
\beqn\label{INF_E_3}
\lim_{N \to \infty \atop J=1,2,3} \frac{E}{3N} = \varepsilon^{GP}.
\eeqn
Furthermore,
since 
$\lim_{N \to \infty \atop J=1,2,3} \beta = 0$
and
$\lim_{N \to \infty \atop J=1,2,3} C_{N,0,0} = 0$,
and, consequently,\break\hfill
$\lim_{N \to \infty \atop J=1,2,3} C_{1,0,0} = 0$
hold,
the reduced one-particle density matrix and the density of species $1$ satisfy
\beqn\label{INF_1_RDMs_3}
\!\!\!\!\!\!\!\!
\!\!\!\!
\lim_{N \to \infty \atop J=1,2,3}
\frac{\rho^{(1)}_1(\x,\x')}{N} =
\phi^{GP}_1(\x) \left\{\phi^{GP}_1(\x')\right\}^\ast, \qquad
\lim_{N \to \infty \atop J=1,2,3}
\frac{\rho^{(1)}_1(\x)}{N} =
\left|\phi^{GP}_1(\x)\right|^2,
\eeqn
thus generalizing literature results for single-species bosons \cite{INF1,INF2,INF_LENZ_2017}
and two-species \cite{HIM_MIX_RDM,BB1,BB2,BB3} and multiple-species bosonic mixtures \cite{BB4},
to a three-species bosonic mixture with an unbound infinite-range interparticle interaction.
Note that since the assignment of the labeling $1$, $2$, and $3$ to the species is arbitrary and, in any case,
the mixture is balanced,
relations (\ref{INF_1_RDMs_3}) equally hold for species $2$ and $3$ in the mixture.

Despite relation (\ref{INF_E_3}) for the energy per particle,
there are nonetheless correlations in the mixture at the infinite-particle-number limit
which are expressed by various quantities.
Let us begin to quantify them. 
The correlation energy is defined as the (positive) difference between the mean-field and many-body energies and reads
\beqn\label{E_cor_3}
& &
E_{cor}=E^{GP}-E=
\frac{3}{2} \Bigg[3N\sqrt{\omega^2 + 2\left(\Lambda_1+2\Lambda_{12}\right)} - \nonumber \\
& &
- 3(N-1) \sqrt{\omega^2+2\left[\left(1+\frac{1}{N-1}\right)\Lambda_1+2\Lambda_{12}\right]}
- 2\sqrt{\omega^2 + 6\Lambda_{12}} - \omega \Bigg], \
\eeqn
compare to the correlation energy of the two-species mixture (\ref{E_cor_2}).
Clearly, in the absence of interspecies interactions the correlation energy (\ref{E_cor_3}) boils down
to three times the correlation energy of a single species, see \cite{INF_LENZ_2017}.

At the infinite-particle-number limit we obtain
\beqn\label{E_cor_3_INF}
& & \!\!\!\!\!\!\!\!
\lim_{N \to \infty \atop J=1,2,3} E_{cor}=
\frac{3}{2} \Bigg[3\sqrt{\omega^2 + 2\left(\Lambda_1+2\Lambda_{12}\right)} -
\frac{3\Lambda_1}{\sqrt{\omega^2 + 2\left(\Lambda_1+2\Lambda_{12}\right)}} -
2\sqrt{\omega^2 + 6\Lambda_{12}} - \omega \Bigg] = \nonumber \\
& &
= \frac{3}{2} \left[ 3 \Omega_1^{GP} - \frac{3\Lambda_1}{\Omega_1^{GP}} - 2\Omega_{123} - \omega \right], \
\eeqn
where the following relation is used between the many-body and mean-field frequencies of the three-species mixture,
$\Omega_1 = \sqrt{\omega^2+2\left[\left(1+\frac{1}{N-1}\right)\Lambda_1+2\Lambda_{12}\right]}
= \Omega_1^{GP}\sqrt{1+\frac{2\Lambda_1}{(N-1)\{\Omega_1^{GP}\}^2}}$.
Expression (\ref{E_cor_3_INF}) tells us that, even at the infinite-particle-number limit when
the three species are 100\% condensed and the many-body and mean-field energies per particle
coincide, there are correlations in the three-species mixture.

Despite relation (\ref{INF_1_RDMs_3}) for $100\%$ condensation of the species,
there are nevertheless particles residing outside the condensed modes
in the mixture at the infinite-particle-number limit.
To compute the depletion of species $1$,
diagonalization of the reduced one-particle density matrix (\ref{rho_1_BBB_3}) with Mehler's formula \cite{Robinson_1977,Schilling_2013,Atoms_2021},
in three spatial dimensions,
\beqn\label{MEHLER_rho1}
& &
\left[\frac{(1-\rho)s}{(1+\rho)\pi}\right]^{\frac{3}{2}}
e^{-\frac{1}{2}\frac{(1+\rho^2)s}{1-\rho^2}\left(\x^2+{\x'}^2\right)}e^{+\frac{2\rho s}{1-\rho^2}\x\x'} = \nonumber \\
& &
= \sum_{n_1,n_2,n_3=0}^\infty (1-\rho)^3\rho^{n_1+n_2+n_3}
\Phi_{n_1,n_2,n_3}(\x;s) \Phi_{n_1,n_2,n_3}(\x';s), \\
& & \Phi_{n_1,n_2,n_3}(\x;s) =
\frac{1}{\sqrt{2^{n_1+n_2+n_3} n_1! n_2! n_3!}}
\left(\frac{s}{\pi}\right)^{\frac{3}{4}} H_{n_1}(\sqrt{s}x_1) H_{n_2}(\sqrt{s}x_2) H_{n_3}(\sqrt{s}x_3) e^{-\frac{1}{2}s \x^2}, \nonumber \
\eeqn
is required.
Here $s>0$ is the scaling,
$1>\rho \ge 0$ is non-negative
for the reduced one-particle density matrix (and for the Schmidt decomposition of the wavefunction below),
$H_n(x)$ are the Hermite polynomials,
and $\x=(x_1,x_2,x_3)$ and likewise here and hereafter
for the Cartesian components of three-dimensional operators.
Note that the normalization in (\ref{MEHLER_rho1}) is aligned with the normalization of the eigenvalues, i.e.,
$\sum_{n_1,n_2,n_3=0}^\infty (1-\rho)^3\rho^{n_1+n_2+n_3}=1$.

Comparing (\ref{rho_1_BBB_3}) and (\ref{MEHLER_rho1}), 
the depletion of species $1$ is then given by
\beqn\label{1RDMs_DEPLT_1_3_SPECIES}
& & d^{(1)}_1 = N\left[1-\left(1-\rho^{(1)}_1\right)^3\right] =
N\rho^{(1)}_1 \left(3-3\rho^{(1)}_1+\left\{\rho^{(1)}_1\right\}^2\right), \nonumber \\
& &
\rho^{(1)}_1 = \frac{\mathcal{W}-1}{\mathcal{W}+1}, \qquad \qquad
\mathcal{W} = \sqrt{\frac{\alpha}{\alpha+C_{1,0,0}}} = 
\\ & &
= \sqrt{\left\{1 + \frac{1}{N} \left[\frac{1}{3}\left(\frac{\omega}{\Omega_1} +
\frac{2\Omega_{123}}{\Omega_1}\right) - 1 \right]\right\}\left\{1 + \frac{1}{N} \left[\frac{1}{3}\left(\frac{\Omega_1}{\omega} + \frac{2\Omega_1}{\Omega_{123}}\right) - 1 \right]\right\}}. \nonumber \
\eeqn
For completeness,
the scaling of the natural orbitals
is
\beqn\label{1RDMs_Scaling_1_3}
s^{(1)}_1 = \sqrt{\alpha(\alpha+C_{1,0,0})} = \Omega_1
\sqrt{\frac{1+\frac{1}{N}\left[\frac{1}{3}\left(\frac{\omega}{\Omega_1}+\frac{2\Omega_{123}}{\Omega_1}\right)-1\right]}
{1+\frac{1}{N}\left[\frac{1}{3}\left(\frac{\Omega_1}{\omega}+\frac{2\Omega_1}{\Omega_{123}}\right)-1\right]}}.
\eeqn
It is clear from (\ref{1RDMs_DEPLT_1_3_SPECIES}) and (\ref{1RDMs_Scaling_1_3})
that $\lim_{N \to \infty \atop J=1,2,3} \frac{d^{(1)}_1}{N} = 0$ and
$\lim_{N \to \infty \atop J=1,2,3} s^{(1)}_1 = \Omega_1^{GP}$ hold, respectively,
which state the equivalent to (\ref{INF_1_RDMs_3}).

Yet, although the depleted fraction vanishes at the infinite-particle-number limit, the depletion does not.
Indeed, at the limit of an infinite number of particles we obtain the depletion
\beqn\label{1RDMs_DEPLT_1_INF_3_SPECIES}
& &
\lim_{N \to \infty \atop J=1,2,3} d^{(1)}_1 =
\frac{3}{4}\left[\frac{1}{3}\frac{\left(\omega-\Omega_1^{GP}\right)^2}{\omega\Omega_1^{GP}} +
\frac{2}{3}\frac{\left(\Omega_{123}-\Omega_1^{GP}\right)^2}{\Omega_{123}\Omega_1^{GP}}\right]. \
\eeqn
Eq.~(\ref{1RDMs_DEPLT_1_INF_3_SPECIES}) reduces to the single-species infinite-particle-number depletion \cite{INF_LENZ_2017}
in the absence of interaction between the three species.
Furthermore, comparison to the expression for the depletion in the two-species mixture
(\ref{1RDMs_DEPLT_1_INF}),
we observe the growing weight of the
relative contribution to the depletion of the interspecies term,
when increasing the number of species from $1$ to $2$ and to $3$ species.
Later on, in Sec.~\ref{BBB_P_SEC}, we compare the depletions of mixtures with a different number of species $P$
to the depletion of the mixture with the smallest number of species, i.e., $P=2$,
and plot the respective ratios of depletions for the purpose of analysis.

The mean-field wavefunction (\ref{MIX_WAV_GP_3}) is a product state
of the coordinates of all particles in the mixture
(as well as of all the species' Jacoby coordinates).
There is, hence, no coupling between the coordinates of
different species
and no entanglement, per definition, between the species
within mean-field theory.
This is not the case in many-body theory.
The many-body wavefunction (\ref{WF_HIM_3_JAC}) 
exhibits that the center-of-mass coordinates $\X_N$, $\Y_N$, and $\Z_N$ are coupled, and only them.
This implies, e.g., entanglement between species $1$ and species $2$ and $3$.
Let us quantify it.
It suffices to explore
the center-of-masses wavefunction
which takes on the form
\beqn\label{CM_JAC_BBB_3_t_1}
& &
\!\!\!\!\!\!\!\!\!\!\!\!
\Psi_{CMs}(\X_{N},\Y_{N},\Z_{N}) =
\left[\frac{(a-b)^2(a+2b)}{\pi^3}\right]^{\frac{3}{4}}
e^{-\frac{1}{2}a\left(\X^2_{N}+\Y^2_{N}+\Z^2_{N}\right)}
e^{-b\left(\X_{N}\Y_{N}+\X_{N}\Z_{N}+\Y_{N}\Z_{N}\right)}. \
\eeqn
We can already see that entanglement persists at the limit of an infinite number of particles,
despite $100\%$ Bose-Einstein condensation of the three species.
Quite generally, $a$ and $b$ are finite and different than zero at this limit,
\beqn\label{a_b_BBB}
\lim_{N \to \infty \atop J=1,2,3} b = b = \frac{1}{3}\left(\omega-\Omega_{123}\right), \qquad
\lim_{N \to \infty \atop J=1,2,3} a = a = \frac{1}{3}\left(2\Omega_{123}+\omega\right),
\eeqn
implying that in (\ref{CM_JAC_BBB_3_t_1}) the three species are entangled while each of the species is $100\%$ condensed,
see (\ref{INF_1_RDMs_3}).
Recall that $\Omega_{123}=\sqrt{\omega^2 + 6\Lambda_{12}}$ [Eq.~(\ref{CM_3_freq})]
is already expressed as a function of the interspecies interaction parameter
$\Lambda_{12}$ which is held fixed at the infinite-particle-number limit.

To perform the Schmidt decomposition
between the center-of-mass coordinate of species $1$ and species $2$ and $3$,
one first needs to define
auxiliary coordinates using the remaining $P-1$ center-of-mass coordinates (i.e., of species $2$ and $3$),
\beqn\label{CM_JAC_BBB_3_t_2}
& &
\bar \Q_2 = \frac{1}{\sqrt{2}}\left(-\Y_N+\Z_N\right), \qquad \bar \Q_3 = \frac{1}{\sqrt{2}}\left(\Y_N+\Z_N\right), \nonumber \\
& &
\Longrightarrow
\left(\frac{\Omega_{123}^2\omega}{\pi^3}\right)^{\frac{3}{4}}
e^{-\frac{1}{2}a\X^2_{N}}
e^{-\frac{1}{2}(a+b)\bar \Q^2_3}
e^{-\sqrt{2}b\X_{N}\bar \Q_3}
e^{-\frac{1}{2}(a-b)\bar \Q^2_2}. \
\eeqn
The form of the wavefunction in (\ref{CM_JAC_BBB_3_t_2})
is not yet amenable to applying Mehler's transformation,
because the coordinates $\X_{N}$ and $\bar \Q_3$ are not `equivalent'
(the pre-factors in front of them are different).
Hence and additionally,
the squeezed coordinates are defined which leads to
\beqn\label{CM_JAC_BBB_3_t_3}
& &
\widetilde \X_{N} = \left(\frac{a}{a+b}\right)^{\frac{1}{4}} \X_{N}, \qquad \widetilde \Q_3 =  \left(\frac{a+b}{a}\right)^{\frac{1}{4}} \bar \Q_3, \nonumber \\
& &
\Longrightarrow
\left(\frac{\Omega_{123}^2\omega}{\pi^3}\right)^{\frac{3}{4}}
e^{-\frac{1}{2} \sqrt{a(a+b)} \left(\widetilde \X^2_{N} + \widetilde \Q^2_3\right)}
e^{-\sqrt{2}b \widetilde \X_{N} \widetilde \Q_3}
e^{-\frac{1}{2}(a-b)\bar \Q^2_2}. \
\eeqn
Note that these two transformations, Eqs.~(\ref{CM_JAC_BBB_3_t_2}) and (\ref{CM_JAC_BBB_3_t_3}),
are not required for the mixture with $P=2$ species,
and only start from $P=3$ species in the mixture.

Now, we can apply Mehler's formula in three spatial dimensions,
but taking care that the normalization is aligned with the normalization of the square of the eigenvalues,\break\hfill
$\left(\frac{s}{\pi}\right)^{\frac{3}{2}}
e^{-\frac{1}{2}\frac{(1+\rho^2)s}{1-\rho^2}\left(\x^2+\y^2\right)}
e^{+\frac{2\rho s}{1-\rho^2}\x \y}
= \sum_{n_1,n_2,n_3=0}^\infty \left(1-\rho^2\right)^{\frac{3}{2}} \rho^{n_1+n_2+n_3} 
\Phi_{n_1,n_2,n_3}(\x;s) \Phi_{n_1,n_2,n_3}(\y;s)$,
for the Schmidt decomposition of (\ref{CM_JAC_BBB_3_t_3}). 
The final result is
\beqn\label{CM_JAC_BBB_3_SQUEEZE_MEHLER_t1}
& &
\Psi_{CMs}(\widetilde \X_{N},\bar \Q_2,\widetilde \Q_3) = \\
& &
= \left(\frac{\Omega_{123}}{\pi}\right)^{\frac{3}{4}} e^{-\frac{1}{2}\Omega_{123}\bar \Q^2_2} 
\!\! \sum_{n_1,n_2,n_3=0}^\infty \left(1-\rho^2\right)^{\frac{3}{2}} \rho^{n_1+n_2+n_3} 
\Phi_{n_1,n_2,n_3}(\widetilde \X_{N};s) \Phi_{n_1,n_2,n_3}(\widetilde \Q_3;s), \nonumber \
\eeqn
with the explicit parameters
\beqn\label{CM_JAC_BBB_3_SQUEEZE_MEHLER_t2}
& &
s = \sqrt{(a-b)(a+2b)} = \sqrt{\Omega_{123}\omega}, \qquad 
\rho = \frac{\mathcal{W}_{SD}-1}{\mathcal{W}_{SD}+1}, \nonumber \\
& &
\mathcal{W}_{SD} = \left[\frac{\sqrt{a(a+b)}-\sqrt{2}b}{\sqrt{a(a+b)}+\sqrt{2}b}\right]^{\frac{1}{2}}
= \left[\frac{\sqrt{1+\frac{2}{9}\frac{\left(\omega-\Omega_{123}\right)^2}{\omega\Omega_{123}}}-\frac{\sqrt{2}}{3}
\frac{\left(\omega-\Omega_{123}\right)}{\sqrt{\omega\Omega_{123}}}}
{\sqrt{1+\frac{2}{9}\frac{\left(\omega-\Omega_{123}\right)^2}{\omega\Omega_{123}}}+
\frac{\sqrt{2}}{3}\frac{\left(\omega-\Omega_{123}\right)}{\sqrt{\omega\Omega_{123}}}}\right]^{\frac{1}{2}}, \
\eeqn
which holds for attractive interspecies interactions,
where $\Omega_{123} > \omega$ and hence 
$b=\frac{1}{3}\left(\omega-\Omega_{123}\right)$ is negative.
For repulsive interspecies interactions,
where $b$ is positive since $\Omega_{123} < \omega$,
just take, say, $\widetilde \Q_3 \to - \widetilde \Q_3$
in (\ref{CM_JAC_BBB_3_t_3},\ref{CM_JAC_BBB_3_SQUEEZE_MEHLER_t1})
and
$\rho \to - \rho$ in (\ref{CM_JAC_BBB_3_SQUEEZE_MEHLER_t2}).
As explicitly seen from (\ref{CM_JAC_BBB_3_SQUEEZE_MEHLER_t2}), 
the Schmidt parameter $\rho$ vanishes
when the coupling $b$ between species $1$ and species $2$ and $3$ does,
as it should be when species $1$ becomes disentangled from species $2$ and $3$.
Comparison the three-species to the two-species [Eq.~(\ref{CM_JAC_BB_2_SQUEEZE_MEHLER_t2})] case shows a more intricate expression.
We shall return to that when dealing with the general $P$-species mixture.

Manifestation of the entanglement between species $1$ and species $2$ and $3$
in an observable is, e.g., the position--momentum uncertainty product, i.e.,
deviations from the uncertainty product of the mean-field separable solution reflect
the above-mentioned
entanglement between the three species.
Inverting relation (\ref{CM_3_vecs}) between the center-of-mass coordinates and their Jacoby coordinates,
one has
\beqn
& & {\Delta^2}_{\hat \X_N} =
\frac{1}{6}\left(\frac{1}{\omega}+\frac{2}{\Omega_{123}}\right)\!\bo, \qquad
{\Delta^2}_{\hat \P_{\X_N}} =
\frac{1}{6}\left(\omega+2\Omega_{123}\right)\!\bo, \
\eeqn
and therefore
\beqn\label{UP_BBB_3}
& &
\!\!\!\!\!\!\!\!
{\Delta^2}_{\hat \X_{CM}} {\Delta^2}_{\hat \P_{\X_{CM}}} =
{\Delta^2}_{\hat \X_N} {\Delta^2}_{\hat \P_{\X_N}} =
\left[1+\frac{2}{9}\frac{\left(\omega-\Omega_{12}\right)^2}{\omega\Omega_{12}}\right]\frac{1}{4}\bo, \
\eeqn
where $\hat \X_{CM} = \frac{\sum_{j=1}^{N} \hat \x_j}{N}$ and
$\hat \P_{\X_{CM}} = \sum_{j=1}^{N} \hat \p_{\x_j}$,
the commutation relations
$\left[\hat \X_N,\hat \P_{\X_N}\right]=\left[\hat \X_{CM},\hat \P_{\X_{CM}}\right]=\bi$,
and $\bo$ is a shorthand notation for $1$ in each of the three Cartesian coordinates and similarly $\bi$ is for $i$.
Finally and for comparison,
\beqn\label{UP_BBB_3_GP}
& &
\left\{{\Delta^2}_{\hat \X_{CM}} {\Delta^2}_{\hat \P_{\X_{CM}}}\right\}^{GP} =
\frac{1}{4}\bo,
\eeqn
as is evident from the interaction-dressed Gaussian-shaped mean-field solution (\ref{MIX_GP_OR_3}). 
The resemblance of expressions for the entanglement (\ref{CM_JAC_BBB_3_SQUEEZE_MEHLER_t1},\ref{CM_JAC_BBB_3_SQUEEZE_MEHLER_t2}) and uncertainty product (\ref{UP_BBB_3})
suggests a deeper connection between them.
We delve into that within the general solution of the $P$-species mixture in the next section. 

So far,
we discussed the properties of the balanced three-species mixture of trapped Bose-Einstein condensates.
We built explicitly and analytically the reduced one-particle density matrices,
and, among others, proved $100\%$ condensation at the limit of an infinite number of particles,
and the nonetheless existence of entanglement between the species in this limit and its expression
in the position--momentum uncertainty product.
All of which generalizes results and properties of the two-species mixture.
We also started to build the additional tools
that are needed to treat more than two species in the mixture.
We now turn to the generalization for the $P$-species mixture
and, afterwards, the investigation of the dependence of properties at the infinite-particle-number limit
on the number of species in the mixture.

\section{Generalization to a mixture of Bose-Einstein condensates with $P$ species}\label{BBB_P_SEC}

With the detailed derivation in the previous section,
we may follow a more concise one below.
Consider $P$ bosonic species and a balanced mixture.
The relative Jacoby coordinates are denoted by
$\X_{J,j}, J=1,\ldots P, j=1,\ldots,N-1$,
and the intraspecies center-of-mass coordinates by
$\X_{J,N}, J=1,\ldots P$.

The Hamiltonian in the species' Jacoby coordinates can be expressed like in Sec.~\ref{BBB_3_SEC}
as a sun of two terms, $\hat H = \hat H_{rels} + \hat H_{CMs}$.
The Hamiltonian of the relative motions is
\beqn\label{HAM_MIX_P_XYZWU_rels}
& &
\!\!\!\!\!\!\!\!\!\!
\hat H_{rels}(\X_{1,1},\ldots,\X_{1,N-1},\ldots,\X_{P,1},\ldots,\X_{P,N-1}) =
\sum_{J=1}^P \sum_{j=1}^{N-1} 
\left( -\frac{1}{2} \frac{\partial^2}{\partial \X_{J,j}^2} + \frac{1}{2} \Omega_1^2 \X_{J,j}^2 \right), \
\eeqn
along with the dressed intraspecies frequencies which are all equal in the balanced mixture,
\beqn
& & \Omega_1 =
\sqrt{\omega^2+2\left[\left(1+\frac{1}{N-1}\right)\Lambda_1+(P-1) \Lambda_{12}\right]}. \
\eeqn
The center-of-masses Hamiltonian is
\beqn\label{HAM_MIX_P_XYZWU_CMs}
& &
\hat H_{CMs}(\X_{1,N},\ldots,\X_{P,N}) =
- \frac{1}{2} \sum_{J=1}^P \frac{\partial^2}{\partial \X_{J,N}^2} +\frac{1}{2} 
\begin{pmatrix}
\X_{1,N} &
\cdots &
\X_{P,N} \cr
\end{pmatrix}
\underline{\underline{\bf O}}
\begin{pmatrix}
\X_{1,N} \cr
\vdots \cr
\X_{P,N} \cr
\end{pmatrix}, \nonumber \\
& &
\underline{\underline{\bf O}} = \left(\omega^2 + 2P\Lambda_{12}\right) \underline{\underline{\bf I}}
-2\Lambda_{12}\underline{\underline{\bf 1}}. \
\eeqn
Diagonalizing the frequencies' matrix $\underline{\underline{\bf O}}$
one finds the eigenvalues and eigenvectors emanating
from the center-of-mass degrees-of-freedom.
There are now $P-1$ equal relative-coordinate frequencies in addition to the center-of-mass frequency,
\beqn\label{CM_P_freq}
& & \Omega_{12\ldots P} = \sqrt{\omega^2 + 2P\Lambda_{12}} \mathrm{\ \ [(P-1) \ roots]}, \quad \omega, \
\eeqn
where the corresponding eigenvectors are nothing but the set of Jacoby coordinates
constructed from the center-of-mass coordinates $\left\{\X_{J,N}\right\}$ of the $P$ species,
\beqn\label{CM_P_vecs}
& &
\!\!\!\!\!\!\!\!
\Q_K = \frac{1}{\sqrt{K(K+1)}} \sum_{J=1}^{K} (\X_{K+1,N}-\X_{J,N}),
\ \ 1 \le K \le P-1,
\quad \Q_{P} = \frac{1}{\sqrt{P}} \sum_{J=1}^{P} \X_{J,N}. \
\eeqn

The ground-state energy of the $P$-species mixture is
\beqn\label{E_HIM_P}
& & E = \frac{3}{2} \left[P(N-1)\Omega_1 + (P-1) \Omega_{12\ldots P} + \omega \right] = \\
& &
= \frac{3}{2}
\left[P (N-1) \sqrt{\omega^2+2\left[\left(1+\frac{1}{N-1}\right)\Lambda_1+(P-1) \Lambda_{12}\right]} +
(P-1)\sqrt{\omega^2 + 2P\Lambda_{12}} + \omega \right], \nonumber \
\eeqn
which generalizes Eq.~(\ref{E_HIM_3}) to the case of $P$ species.
Of course, all frequencies must be positive for the $P$-species mixture to be bound, hence
\beqn\label{FREQ_P_BOUNDS}
\Lambda_{12} > - \frac{\omega^2}{2P}, \qquad
\Lambda_1 > - \left(1-\frac{1}{N}\right)\left[\frac{\omega^2}{2} + (P-1)\Lambda_{12}\right]. \
\eeqn
As a result,
the energy (\ref{E_HIM_P}) is bound from below but not from above.
One can now investigate the dependence of the correlation energy on the number of species,
see below and in the next section.

We proceed straightforwardly 
to the all-particle density matrix expressed using the 
Jacoby coordinates of the individual $P$ species,
\beqn\label{PN_DENS_JAC}
& & \Psi(\X_{1,1},\ldots,\X_{1,N},\ldots,\X_{1,P},\ldots,\X_{P,N})
\Psi^\ast(\X'_{1,1},\ldots,\X'_{1,N},\ldots,\X'_{1,P},\ldots,\X'_{P,N}) = \nonumber \\
& & =
\left(\frac{\Omega_1}{\pi}\right)^{\frac{3P(N-1)}{2}}
\left(\frac{\Omega_{12\ldots P}}{\pi}\right)^{\frac{3(P-1)}{2}}
\left(\frac{\omega}{\pi}\right)^{\frac{3}{2}} \,
e^{-\frac{1}{2} \Omega_1 \sum_{J=1}^P\sum_{j=1}^{N-1} \left(\X_{J,j}^2+{\X'}_{J,j}^2\right)} \times \nonumber \\
& & \times
e^{-\frac{1}{2} a \sum_{J=1}^P \left(\X_{J,N}^2+{\X'}_{J,N}^2\right)}
e^{-b \sum_{1=J<K}^P \left(\X_{J,N}\X_{K,N}+\X'_{J,N}\X'_{K,N}\right)}, \
\eeqn
where the coefficients are
\beqn\label{ab_coeff_BBBB}
& &
a = \frac{1}{P}\left[(P-1)\Omega_{12\ldots P}+\omega\right]=\Omega_{12\ldots P}+b, \qquad
b = \frac{1}{P}\left(\omega-\Omega_{12\ldots P}\right). \
\eeqn
These coefficients are interrelated and obey
$a+(P-1)b=\omega$.

The reduced one-particle density matrix of species $1$ is therefore given by the following integration procedure.
To this end, we need to integrate first over the relative coordinates of species $P$, $P-1$, and so on, down to species $2$.
Then, we need to integrate over the corresponding
center-of-mass coordinates $\X'_{P,N}=\X_{P,N}$, $\X'_{P-1,N}=\X_{P-1,N}$, and so on,
down to $\X'_{2,N}=\X_{2,N}$.
The first of these center-of-mass integrations, over $\X_{P,N}$ of species $P$,
renormalizes the coefficients in the reduced center-of-masses density matrix
of the $P-1$ species preceding species $P$.
The second integration, over $\X_{P-1,N}$ of species $P-1$,
renormalizes the coefficients in the reduced center-of-masses density matrix
of the $P-2$ species preceding species $P-1$, and so on.
This procedure defines a recurrence relation that can be solved using mathematical induction,
see appendix \ref{SEC_FOLD}.
All in all,
after some algebra
one arrives at the following, final and compact expression
\beqn\label{C_N_0_0_0_P}
C_{N,0,\ldots,0} = - \frac{1}{N}\frac{(P-1)b^2}{a+(P-2)b} =
- \frac{1}{N}\frac{P-1}{P} \frac{\left(\omega-\Omega_{12\ldots P}\right)^2}{(P-1)\omega+\Omega_{12\ldots P}}.
\eeqn
It is quite appealing that folding down
the effect of $P-1$ center-of-mass coordinates
onto species $1$ can be performed analytically and moreover
leads to the exact and compact expression (\ref{C_N_0_0_0_P}).
The results for the specific cases of $P=2,3$ species are readily recovered,
as well as for the single-species case ($P=1$),
where,
in accordance with Ref.~\cite{HIM_Cohen}, $C_{N}=0$.

The working expression for 
the reduced all-particle density matrix of species $1$
in the $P$-species mixture is
\beqn\label{F_1_0_0_BBBB_P}
& & e^{-\frac{\alpha}{2} \sum_{j=1}^{N} \left(\x_j^2+{\x'}_j^2\right) - \beta \sum_{1\le j <k}^{N}
\left(\x_j \x_k + \x'_j \x'_k\right)} e^{-\frac{1}{4}C_{N,0,\ldots,0}\left\{\sum_{j=1}^{N}\left(\x_j+\x'_j\right)\right\}^2}, \
\eeqn
with
\beqn\label{F_1_0_0_BBBB_P_Coeff_1}
& & \alpha = \Omega_1 + \frac{1}{N}\left(a-\Omega_1\right)
= \Omega_1\left(1+\frac{1}{N}\left\{\frac{1}{P}\left[\frac{\omega}{\Omega_1}+\frac{(P-1)\Omega_{12\ldots P}}{\Omega_1}\right]-1\right\}\right), \nonumber \\
& &
\beta = \alpha-\Omega_1 =
\Omega_1\frac{1}{N}\left\{\frac{1}{P}\left[\frac{\omega}{\Omega_1}+\frac{(P-1)\Omega_{12\ldots P}}{\Omega_1}\right]-1\right\}, \
\eeqn
thus generalizing the $P=1,2,3$ results.

The reduced one-particle density matrix and the density of species $1$ are
then given by
\beqn\label{rho_1_BBBB_P}
& & \rho^{(1)}_1(\x,\x') = N \left(\frac{\alpha+C_{1,0,\ldots,0}}{\pi}\right)^{\frac{3}{2}}
e^{-\frac{\alpha}{2}\left(\x^2+{\x'}^2\right)}e^{-\frac{1}{4}C_{1,0,\ldots,0}\left(\x+\x'\right)^2}, \nonumber \\
& & \rho^{(1)}_1(\x) = N \left(\frac{\alpha+C_{1,0,\ldots,0}}{\pi}\right)^{\frac{3}{2}}
e^{-\left(\alpha+C_{1,0,\ldots,0}\right)\x^2}, \
\eeqn
where
\beqn\label{rho_1_BBBB_P_coeff}
& &
\alpha + C_{1,0,\ldots,0} = \left(\alpha-\beta\right) \frac{\left(\alpha-\beta\right) + N\left(C_{N,0,\ldots,0}+\beta\right)}
{\left(\alpha-\beta\right) + (N-1)\left(C_{N,0,\ldots,0}+\beta\right)} = 
\frac{1}{1+\frac{1}{N}\left\{\frac{1}{P}\left[\frac{\Omega_1}{\omega}+\frac{(P-1)\Omega_1}{\Omega_{12\ldots P}}\right] - 1 \right\}}. \nonumber \\ \
\eeqn
Here,
$C_{N,0,\ldots,0} + \beta =
\frac{1}{N}\left[\frac{P\omega\Omega_{12\ldots P}}{(P-1)\omega+\Omega_{12\ldots P}} - \Omega_1\right]
= - \frac{P\omega\Omega_{12\ldots P}}{(P-1)\omega+\Omega_{12\ldots P}} \frac{1}{N}
\left\{\frac{1}{P} \left[\frac{\Omega_1}{\omega} + \frac{(P-1)\Omega_1}{\Omega_{12\ldots P}}\right] - 1\right\}$
and
the auxiliary relation
$C_{N,0,\ldots,0} + \alpha = \Omega_1\left(1-\frac{1}{N}\right)+\frac{1}{N}\left[\frac{(a-b)[a+(P-1)b]}{a+(P-2)b}\right]
= \Omega_1\left(1-\frac{1}{N}\right)+\frac{1}{N}\left[\frac{P\omega\Omega_{12\ldots P}}{(P-1)\omega+\Omega_{12\ldots P}}\right]$
are utilized.
We emphasize that expressions (\ref{rho_1_BBBB_P},\ref{rho_1_BBBB_P_coeff}) are exact and explicitly given quantities,
where all the interactions with and impact of species $2,\ldots,P$
are exactly folded onto
species $1$.

Diagonalizing the reduced one-particle density matrix (\ref{rho_1_BBBB_P},\ref{rho_1_BBBB_P_coeff})
using Mehler's formula \cite{Robinson_1977,Schilling_2013,Atoms_2021}, see (\ref{MEHLER_rho1}),
the depletion is given by
\beqn\label{1RDMs_DEPLT_1_P_SPECIES}
& & d^{(1)}_1 = N\left[1-\left(1-\rho^{(1)}_1\right)^3\right] =
N\rho^{(1)}_1 \left(3-3\rho^{(1)}_1+\left\{\rho^{(1)}_1\right\}^2\right), \nonumber \\
& &
\rho^{(1)}_1 = \frac{\mathcal{W}-1}{\mathcal{W}+1}, \qquad \qquad
\mathcal{W} = \sqrt{\frac{\alpha}{\alpha+C_{1,0,\ldots,0}}} = \\
& &
= \sqrt{\left\{1 + \frac{1}{N} \left[\frac{1}{P}\left(\frac{\omega}{\Omega_1} +
\frac{(P-1)\Omega_{12\ldots P}}{\Omega_1}\right) - 1 \right]\right\}\left\{1 + \frac{1}{N} \left[\frac{1}{P}\left(\frac{\Omega_1}{\omega} + \frac{(P-1)\Omega_1}{\Omega_{12\ldots P}}\right) - 1 \right]\right\}}, \nonumber \
\eeqn
and boils down to the specific cases for $P=1,2,3$ species discussed before,
in \cite{INF_LENZ_2017}, appendix \ref{SEC_MIX2}, and in the previous section, respectively.

To characterize the entanglement between species $1$ and the remaining $P-1$ species,
Schmidt decomposition of the $P$-species center-of-masses wavefunction is required.
Thus, we write
\beqn\label{CM_JAC_BBBB_P_t}
& &
\Psi_{CMs}(\X_{1,N},\ldots,\X_{P,N}) =
\left[\frac{(a-b)^{P-1}\left[a+(P-1)b\right]}{\pi^P}\right]^{\frac{3}{4}}
e^{-\frac{1}{2} a \sum_{J=1}^P \X_{J,N}^2}
e^{-b \sum_{1=J<K}^P \X_{J,N}\X_{K,N}} = \nonumber \\
& &
= \left(\frac{\Omega_{12\ldots P}^{P-1}\omega}{\pi^P}\right)^{\frac{3}{4}}
e^{-\frac{1}{2}a\X^2_{1,N}}
e^{-\frac{1}{2}\left[a+(P-2)b\right]\bar \Q^2_P}
e^{-\sqrt{(P-1)}b\X_{1,N}\bar \Q_P}
e^{-\frac{1}{2} (a-b) \sum_{K=2}^{P-1} \bar \Q_K^2}
= \nonumber \\
& &
= \left(\frac{\Omega_{12\ldots P}^{P-1}\omega}{\pi^P}\right)^{\frac{3}{4}}
e^{-\frac{1}{2} \sqrt{a\left[a+(P-2)b\right]} \left(\widetilde \X^2_{1,N} + \widetilde \Q^2_P\right)}
e^{-\sqrt{(P-1)}b \widetilde \X_{1,N} \widetilde \Q_P}
e^{-\frac{1}{2} (a-b) \sum_{K=2}^{P-1} \bar \Q_K^2} = \\
& &
= \left(\frac{\Omega_{12\ldots P}}{\pi}\right)^{\frac{3(P-2)}{4}} e^{-\frac{1}{2}\Omega_{12\ldots P} \sum_{K=2}^{P-1} \bar \Q_K^2} 
\times \nonumber \\
& &
\times \sum_{n_1,n_2,n_3=0}^\infty \left(1-\rho^2\right)^{\frac{3}{2}} \rho^{n_1+n_2+n_3} 
\Phi_{n_1,n_2,n_3}(\widetilde \X_{1,N};s) \Phi_{n_1,n_2,n_3}(\widetilde \Q_P;s), \nonumber \
\eeqn
where one needs to define auxiliary coordinates using the remaining $P-1$ center-of-mass coordinates (i.e., of species $2,\ldots,P$),
\beqn\label{CM_AUX_vecs_2_3_P}
& &
\!\!\!\!
\bar \Q_K = \frac{1}{\sqrt{(K-1)K}} \sum_{J=2}^{K} (\X_{K+1,N}-\X_{J,N}),
\ \ 2 \le K \le P-1, \ \quad \bar \Q_P = \frac{1}{\sqrt{P-1}} \sum_{J=2}^P \X_{J,N}, \nonumber \\ \
\eeqn
and, furthermore, the squeezed coordinates
\beqn\label{CM_SQUEEZE_vecs_1_P}
\widetilde \X_{1,N} = \left[\frac{a}{a+(P-2)b}\right]^{\frac{1}{4}} \X_{1,N},
\qquad
\widetilde \Q_P = \left[\frac{a+(P-2)b}{a}\right]^{\frac{1}{4}} \bar \Q_P. \
\eeqn
Recall that these two transformations are not required for the mixture with $P=2$ species,
as is evident
from the ranges of the summations in (\ref{CM_AUX_vecs_2_3_P}) and
by substituting $P=2$ in (\ref{CM_SQUEEZE_vecs_1_P}).
In obtaining (\ref{CM_JAC_BBBB_P_t}) we also made use of the relation
$\sum_{2=J<K}^P \X_{J,N}\X_{K,N} = \frac{1}{2} \left[(P-2)\bar \Q_P^2 - \sum_{K=2}^{P-1} \bar \Q_K^2 \right]$
connecting a set of $P-1$ coordinates and the Jacoby coordinates constructed from them.

The solution for the Schmidt decomposition
is then given by
\beqn\label{CM_JAC_BBBB_P_SQUEEZE_MEHLER_t2}
& &
s = \sqrt{(a-b)\left[a+(P-1)b\right]} = \sqrt{\Omega_{12\ldots P}\omega}, \qquad \qquad
\rho = \frac{\mathcal{W}_{SD}-1}{\mathcal{W}_{SD}+1}, \\
& &
\mathcal{W}_{SD} = \left[\frac{\sqrt{a\left[a+(P-2)b\right]}-\sqrt{(P-1)}b}{\sqrt{a\left[a+(P-2)b\right]}+\sqrt{(P-1)}b}\right]^{\frac{1}{2}}
= \left[\frac{\sqrt{1+\frac{P-1}{P^2}\frac{\left(\omega-\Omega_{12\ldots P}\right)^2}{\omega\Omega_{12\ldots P}}}-\frac{\sqrt{P-1}}{P}
\frac{\left(\omega-\Omega_{12\ldots P}\right)}{\sqrt{\omega\Omega_{12\ldots P}}}}
{\sqrt{1+\frac{P-1}{P^2}\frac{\left(\omega-\Omega_{12\ldots P}\right)^2}{\omega\Omega_{12\ldots P}}}+\frac{\sqrt{P-1}}{P}
\frac{\left(\omega-\Omega_{12\ldots P}\right)}{\sqrt{\omega\Omega_{12\ldots P}}}}\right]^{\frac{1}{2}}, \nonumber \
\eeqn
which holds for attractive interspecies interactions ($b<0$).
For repulsive interspecies interactions ($b>0$),
merely take, say, $\widetilde \Q_P \to - \widetilde \Q_P$ in (\ref{CM_JAC_BBBB_P_t},\ref{CM_AUX_vecs_2_3_P})
and
$\rho \to - \rho$ in (\ref{CM_JAC_BBBB_P_SQUEEZE_MEHLER_t2}).
The explicit results for $P=2$ discussed before \cite{Atoms_2021}, also see appendix \ref{SEC_MIX2},
and for $P=3$ presented in Sec.~\ref{BBB_3_SEC}
emerge from the general expression for $P$ species (\ref{CM_JAC_BBBB_P_SQUEEZE_MEHLER_t2}).

Collecting the eigenvalues squares in the Schmidt decomposition (\ref{CM_JAC_BBBB_P_t}),
$\left(1-\rho^2\right)^3 \rho^{2(n_1+n_2+n_3)}$, $n_1,n_2,n_3=0,1,2,3,\ldots$,
one gets for the von Neumann entanglement entropy
\beqn\label{S_vNEE_RHO}
& &
{\mathcal S} = - \sum_{n_1,n_2,n_3=0}^\infty \left(1-\rho^2\right)^3 \rho^{2(n_1+n_2+n_3)}
\ln\left[\left(1-\rho^2\right)^3 \rho^{2(n_1+n_2+n_3)}\right] = \nonumber \\
& &
= -3\left[\ln\left(1-\rho^2\right) + \frac{\rho^2\ln\left(\rho^2\right)}{1-\rho^2}\right]  \
\eeqn
between speceis $1$ and the other $P-1$ species,
which holds true also at the infinite-particle-number limit.
The von Neumann entanglement entropy ${\mathcal S}$ is a
monotonously increasing function of the Schmidt-decomposition parameter $\rho^2$,
and we hence refer to either of the quantities interchangeably. 

The mean-field solution of the $P$-species mixture goes as follows.
The ansatz for the wavefunction is the separable product state
\beq\label{MIX_WAV_GP_P}
\Phi^{GP}(\x_{1,1},\ldots,\x_{1,N},\ldots,\x_{P,1},\ldots,\x_{P,N}) = 
\prod_{J=1}^P\prod_{j=1}^{N} \phi^{GP}_J(\x_{J,j}).
\eeq
We note that no prior assumption is made on the shapes of the orbitals of the different species.
Sandwiching the Hamiltonian $\hat H = \hat H_{CMs} + \hat H_{rels}$ with the ansatz (\ref{MIX_WAV_GP_P}),
minimizing the resulting energy functional with respect to
the shapes of the orbitals $\phi^{GP}_J(\x), J=1,\ldots,P$,
the $P$-coupled Gross-Pitaevskii equations of the $P$-species mixture are derived,
\beqn\label{MIX_EQ_GP_BBB_P}
& &
\!\!
\left\{ -\frac{1}{2} \frac{\partial^2}{\partial \x^2} + \frac{1}{2} \omega^2 \x^2 
+  \int d\x' \left[\Lambda_1 |\phi_J(\x')|^2 + \Lambda_{12} \sum_{K\ne J}^P |\phi_K(\x')|^2\right]\!(\x-\x')^2\right\} \phi_J(\x) =
\mu_J\phi_J(\x), \nonumber \\
& & J=1,\ldots P,
\eeqn
where $\mu_J, J=1,\ldots P$ are the chemical potentials of the species
(or, equivalently, the respective Lagrange multipliers ensuring the normalization of the orbitals to unity).
The solution for the balanced multiple-species mixture is simply
\beqn\label{MIX_GP_OR_P}
& & \phi^{GP}_J(\x) = \left(\frac{\Omega_1^{GP}}{\pi}\right)^{\frac{3}{4}}
e^{-\frac{1}{2}\Omega_1^{GP}\x^2}, \qquad J=1,\ldots,P, \nonumber \\
& &
\Omega_1^{GP} = \sqrt{\omega^2 + 2\left[\Lambda_1+(P-1)\Lambda_{12}\right]}, \
\eeqn
where $\mu_J=\frac{3}{2}\left[\Omega_1^{GP}+\frac{\Lambda_1+(P-1)\Lambda_{12}}{\Omega_1^{GP}}\right]$.

The Gross-Pitaevskii energy per particle
of the $P$-species mixture reads
\beqn\label{MIX_GP_E_N_P}
& & \!\!\!\!\!\!\!\!\!\!\!\!
\varepsilon^{GP} = 
\frac{E^{GP}}{PN} =
\frac{3}{2}\Omega_1^{GP}=
\frac{3}{2}\sqrt{\omega^2 + 2\left[\Lambda_1+(P-1)\Lambda_{12}\right]}.
\eeqn
We can now proceed to characterize correlations at the infinite-particle-number limit.

But first,
the infinite-particle-number limit implies $100\%$ Bose-Einstein condensation and boiling down
to some of the mean-field quantities,
which needs to be proved.
Consequently, one finds for the energy of the multiple-species mixture
\beqn\label{INF_E_P}
\lim_{N \to \infty \atop J=1,\ldots,P} \frac{E}{PN} = \varepsilon^{GP}.
\eeqn
Furthermore, for the reduced one-particle density matrix and the density of species $1$ one gets
\beqn\label{INF_1_RDMs_P}
\!\!\!\!\!\!\!\!
\!\!\!\!
\lim_{N \to \infty \atop J=1,\ldots,P}
\frac{\rho^{(1)}_1(\x,\x')}{N} =
\phi^{GP}_1(\x) \left\{\phi^{GP}_1(\x')\right\}^\ast, \qquad
\lim_{N \to \infty \atop J=1,\ldots,P}
\frac{\rho^{(1)}_1(\x)}{N} =
\left|\phi^{GP}_1(\x)\right|^2,
\eeqn
thus generalizing literature results for
single-species bosons \cite{INF1,INF2,INF_LENZ_2017},
two-species bosonic mixtures \cite{HIM_MIX_RDM,BB1,BB2,BB3},
and multiple-species bosonic mixtures \cite{BB4}, 
for $P$-species bosonic mixtures with an unbound infinite-range interparticle interaction.
Note that since the assignment of the labeling $1,\ldots,P$ to the species is arbitrary and, anyhow, the mixture is balanced,
relations (\ref{INF_1_RDMs_P}) equally hold for species $2,\ldots,P$ in the mixture.
All in all,
these are generalizing the results from the previous section for three-species bosonic mixtures
to bosonic mixtures with an arbitrary number of $P$ species.

The correlation energy reads
\beqn\label{E_cor_P}
& &
E_{cor}=E^{GP}-E=
\frac{3}{2} \Bigg[PN\sqrt{\omega^2 + 2\left[\Lambda_1+(P-1)\Lambda_{12}\right]} - \\
& &
- P(N-1) \sqrt{\omega^2+2\left[\left(1+\frac{1}{N-1}\right)\Lambda_1+(P-1)\Lambda_{12}\right]}
- (P-1)\sqrt{\omega^2 + 2P\Lambda_{12}} - \omega \Bigg], \nonumber \
\eeqn
where (\ref{E_cor_P}) can be
compared to the correlation energy of the three-species mixture (\ref{E_cor_3}).
Clearly, in the absence of interspecies interactions the correlation energy (\ref{E_cor_P}) boils down
to ($P$ times) the correlation energy of a single species.

At the infinite-particle-number limit we thus obtain
\beqn\label{E_cor_P_INF}
& &
\lim_{N \to \infty \atop J=1,\ldots,P} E_{cor}=
\frac{3}{2} \Bigg[P\sqrt{\omega^2 + 2\left[\Lambda_1+(P-1)\Lambda_{12}\right]} -
\frac{P\Lambda_1}{\sqrt{\omega^2 + 2\left[\Lambda_1+(P-1)\Lambda_{12}\right]}} - \nonumber \\
& &
- (P-1)\sqrt{\omega^2 + 2P\Lambda_{12}} - \omega \Bigg]
= \frac{3}{2} \left[ P \Omega_1^{GP} - \frac{P\Lambda_1}{\Omega_1^{GP}} - (P-1)\Omega_{12\ldots P} - \omega \right], \
\eeqn
where
$\Omega_1 = \sqrt{\omega^2+2\left[\left(1+\frac{1}{N-1}\right)\Lambda_1+(P-1)\Lambda_{12}\right]}
= \Omega_1^{GP}\sqrt{1+\frac{2\Lambda_1}{(N-1)\{\Omega_1^{GP}\}^2}}$ is used.
Expression (\ref{E_cor_P_INF}) asserts that,
even at the limit of an infinite number of particles, when
the $P$ species are 100\% condensed and the many-body and mean-field energies per particle coincide,
there are correlations in the $P$-species mixture.

Next, at the infinite-particle-number limit we obtain the depletion
\beqn\label{1RDMs_DEPLT_1_INF_P_SPECIES}
& &
\lim_{N \to \infty \atop J=1,2,\ldots,P} d^{(1)}_1 =
\frac{3}{4}\left[\frac{1}{P}\frac{\left(\omega-\Omega_1^{GP}\right)^2}{\omega\Omega_1^{GP}} +
\frac{P-1}{P}\frac{\left(\Omega_{12\ldots P}-\Omega_1^{GP}\right)^2}{\Omega_{12\ldots P}\Omega_1^{GP}}\right]. \
\eeqn
Eq.~(\ref{1RDMs_DEPLT_1_INF_P_SPECIES}) reduces to the single-species infinite-particle-number depletion \cite{INF_LENZ_2017}
in the absence of interaction between the $P$ species.
Later on, we compare the depletions of mixtures with a different number of species $P$
to that of the `smallest' mixture with $P=2$ species.

Furthermore, entanglement persists at the limit of an infinite number of particles,
despite $100\%$ Bose-Einstein condensation of the $P$ species.
This can be readily seen
when examining the center-of-masses wavefunction (\ref{CM_JAC_BBBB_P_t}).
Explicitly, at the infinite-particle-number limit $a$ and $b$, see Eq.~(\ref{ab_coeff_BBBB}),
remain finite and different than zero,
\beqn\label{a_b_BBBB}
\lim_{N \to \infty \atop J=1,\ldots,P} b = b, \qquad
\lim_{N \to \infty \atop J=1,\ldots,P} a = a,
\eeqn
where $\lim_{N \to \infty \atop J=1,\ldots,P} \Omega_{12\ldots P} = \Omega_{12\ldots P}$ is satisfied, of course.
Consequently,
species $1$ is entangled with the other $P-1$ species, Eq.~(\ref{CM_JAC_BBBB_P_t}),
at the infinite-particle-number limit.

Manifestation of that in an observable is,
for example, the position--momentum uncertainty product.
In other words,
deviations from the mean-field separable-solution uncertainty product reflects
the entanglement between the $P$ species.
Inverting (\ref{CM_P_vecs}) one obtains
\beqn
& & {\Delta^2}_{\hat \X_{1,N}} =
\frac{1}{2P}\left[\frac{1}{\omega}+\frac{P-1}{\Omega_{12\ldots P}}\right]\!\bo,
\qquad
{\Delta^2}_{\hat \P_{\X_{1,N}}} =
\frac{1}{2P}\left[\omega+\left(P-1\right)\Omega_{12\ldots P}\right]\!\bo,
 \
\eeqn
and hence
\beqn\label{UP_BBB_P}
& &
\!\!\!\!\!\!\!\!
{\Delta^2}_{\hat \X_{1,CM}} {\Delta^2}_{\hat \P_{\X_{1,CM}}} =
{\Delta^2}_{\hat \X_{1,N}} {\Delta^2}_{\hat \P_{\X_{1,N}}} =
\left[1+\frac{P-1}{P^2}\frac{\left(\omega-\Omega_{12\ldots P}\right)^2}{\omega\Omega_{12\ldots P}}\right]\frac{1}{4}\bo, \
\eeqn
where $\hat \X_{1,CM} = \frac{\sum_{j=1}^{N} \hat \x_{1,j}}{N}$ and
$\hat \P_{\X_{1,CM}} = \sum_{j=1}^{N} \hat \p_{\x_{1,j}}$.
Finally,
\beqn\label{UP_BBB_P_GP}
& &
\left\{{\Delta^2}_{\hat \X_{1,CM}} {\Delta^2}_{\hat \P_{\X_{1,CM}}}\right\}^{GP} =
\frac{1}{4}\bo,
\eeqn
as is evident from the interaction-dressed Gaussian-shaped mean-field solution (\ref{MIX_GP_OR_P}). 

Furthermore,
it is possible to express the uncertainty product (\ref{UP_BBB_P})
as a function of the Schmidt-decomposition parameter $\rho^2$ and vice versa, see below.
The connection between these two many-particle quantities is appealing.
This can be done by calculating the uncertainty product
directly from the left-hand-side of the Mehler's-formula representation [see above (\ref{CM_JAC_BBB_3_SQUEEZE_MEHLER_t1})]
of the center-of-masses wavefunction (\ref{CM_JAC_BBBB_P_t}),
where
${\Delta^2}_{\hat \X_{1,CM}} {\Delta^2}_{\hat \P_{\X_{1,CM}}}
= {\Delta^2}_{\widetilde \X_{1,N}} {\Delta^2}_{\P_{\widetilde \X_{1,N}}}$
is used.
The final result is compact and intriguing,
\beqn\label{rhoWSD}
& &
{\Delta^2}_{\hat \X_{1,CM}} {\Delta^2}_{\hat \P_{\X_{1,CM}}} = \left(\frac{1+\rho^2}{1-\rho^2}\right)^2 \frac{1}{4}\bo, \
\eeqn
compare to (\ref{UP_BBB_P}).

Last but not least,
inverting (\ref{rhoWSD})
we can express the exact relation between the Schmidt-decomposition parameter
and the uncertainty product as
\beqn\label{WSDrho_1}
& &
\rho = 
\left[\frac{{\Delta}_{\hat X_{1,CM}} {\Delta}_{\hat P_{1,X_{CM}}}-\frac{1}{2}}
{{\Delta}_{\hat X_{1,CM}} {\Delta}_{\hat P_{X_{1,CM}}}+\frac{1}{2}}\right]^{\frac{1}{2}},
\eeqn
which holds for attractive as well as repulsive interspecies interactions.
Here, ${\Delta^2}_{\hat X_{1,CM}} {\Delta^2}_{\hat P_{X_{1,CM}}}$ is the component of the uncertainty product (\ref{UP_BBB_P}) in any of
the Cartesian directions.
Alternatively,
comparing (\ref{CM_JAC_BBBB_P_SQUEEZE_MEHLER_t2}) and (\ref{UP_BBB_P}) directly
leads to another form,
\beqn\label{WSDrho_2}
& &
\rho = \frac{\mathcal{W}_{SD}-1}{\mathcal{W}_{SD}+1}, \qquad
\mathcal{W}_{SD} =
\left[\frac{{\Delta}_{\hat X_{1,CM}} {\Delta}_{\hat P_{X_{1,CM}}}+\sqrt{{\Delta^2}_{\hat X_{1,CM}} {\Delta^2}_{\hat P_{X_{1,CM}}}-\frac{1}{4}}}
{{\Delta}_{\hat X_{1,CM}} {\Delta}_{\hat P_{X_{1,CM}}}-\sqrt{{\Delta^2}_{\hat X_{1,CM}} {\Delta^2}_{\hat P_{X_{1,CM}}}-\frac{1}{4}}}\right]^{\frac{1}{2}}. \
\eeqn
Of course, both approaches are equivalent and expressions (\ref{WSDrho_1}) and (\ref{WSDrho_2})
for the connection between the Schmidt-decomposition parameter and the uncertainty product are equal to each other.

Finally, one can express the von Neumann entanglement entropy (\ref{S_vNEE_RHO}) 
in terms of the uncertainty product,
\beqn\label{S_vNEE_UP}
& &
\!\!\!\!
{\mathcal S} = 3 \left[\ln\left({{\Delta}_{\hat X_{1,CM}} {\Delta}_{\hat P_{X_{1,CM}}}+\frac{1}{2}}\right) +
\left({{\Delta}_{\hat X_{1,CM}} {\Delta}_{\hat P_{X_{1,CM}}}-\frac{1}{2}}\right)
\ln\left(\frac{{\Delta}_{\hat X_{1,CM}} {\Delta}_{\hat P_{X_{1,CM}}}+\frac{1}{2}}
{{\Delta}_{\hat X_{1,CM}} {\Delta}_{\hat P_{X_{1,CM}}}-\frac{1}{2}}\right)\right]. \nonumber \\ \
\eeqn
The connection (\ref{S_vNEE_UP}) is simple and interesting.
All in all, owing to (\ref{a_b_BBBB}),
the above results hold at the infinite-particle-number limit of the $P$-species mixtures.
We can proceed to investigate their properties.

\section{Correlations at the infinite-particle-number limit and their dependence
on the number of species $P$ in the mixture}\label{SEC_INVESTIGATION}

We now move to apply the above-derived theory at the infinite-particle-number limit of a multiple-species bosonic mixture.
We have shown that at this limit the species are 100\% condensed,
and that the many-body energy per particle and densities per particle 
coincide with their mean-field counterparts.
The main questions to be addressed
are which correlations do exist in a mixture,
how they depend on the interactions,
and, in particular, on the number of species in the mixture
at the infinite-particle-number limit?

\begin{figure}[!]
\begin{center}
\hglue -1.6 truecm
\includegraphics[width=0.43\columnwidth,angle=-90]{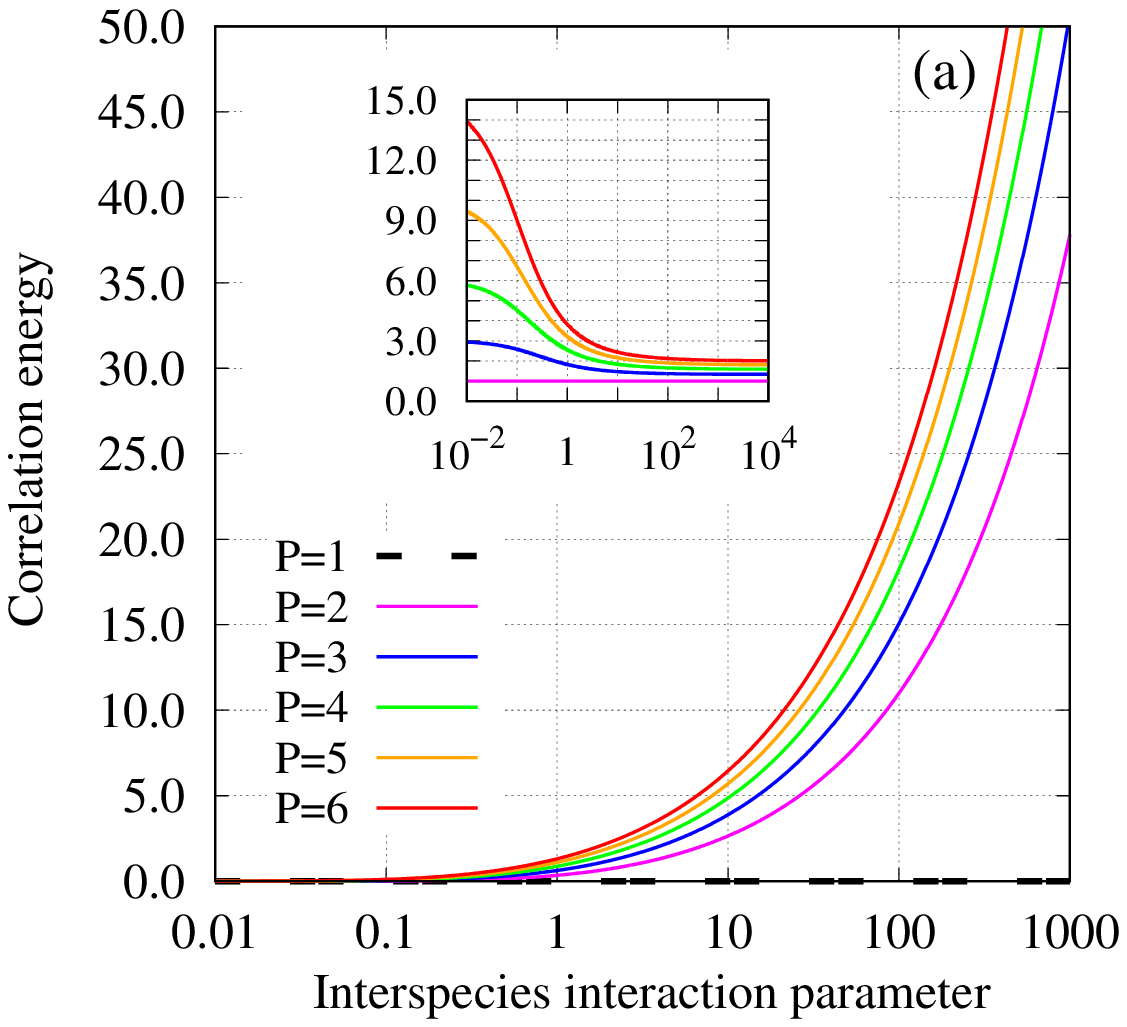}
\hglue -2.4 truecm
\includegraphics[width=0.43\columnwidth,angle=-90]{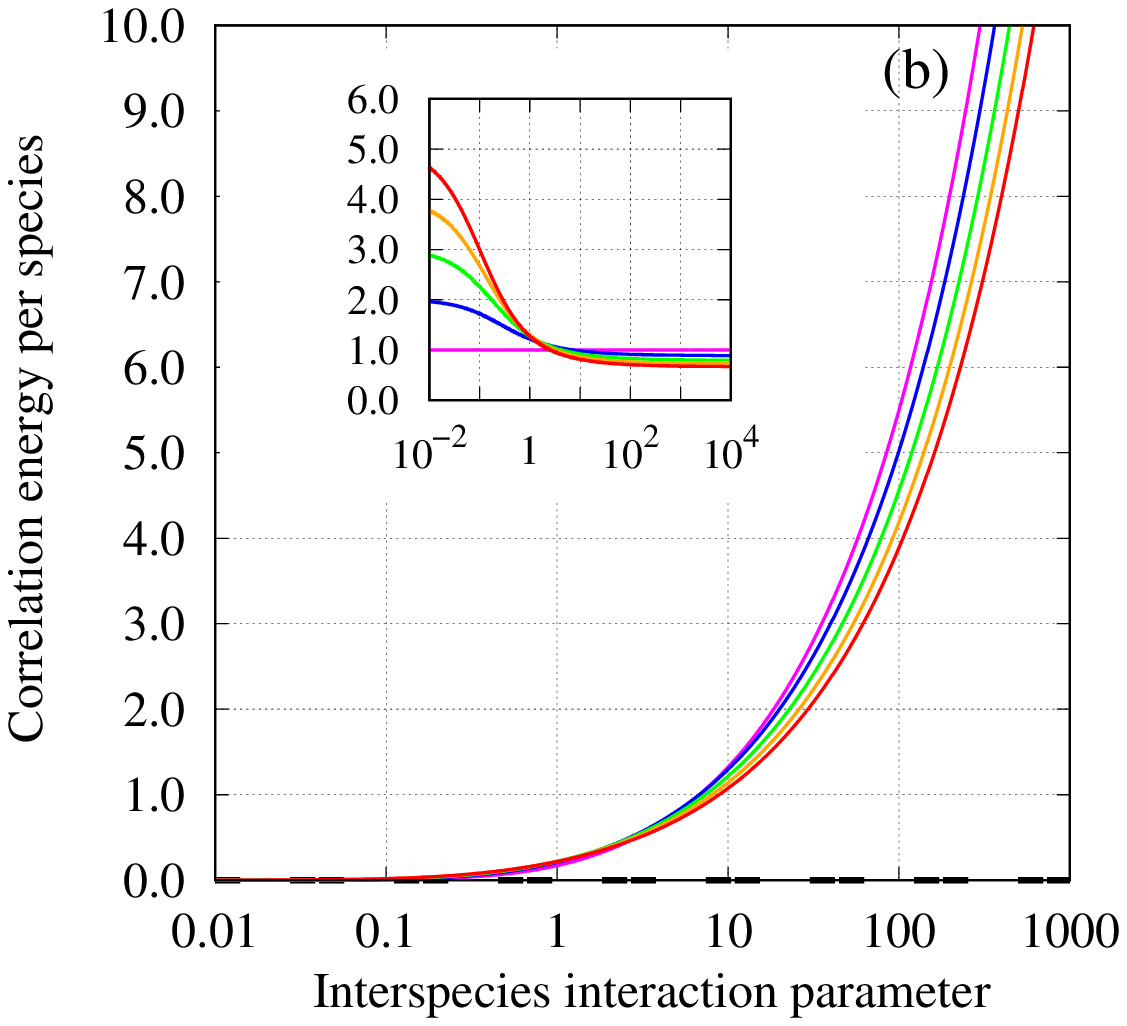}
\end{center}
\vglue 0.50 truecm
\caption{(a) Correlation energy, $E_{cor}$, at the infinite-particle-number limit
of $P$ species of bosons as a function of the interspecies interaction parameter $\Lambda_{12}$.
The intraspecies interactions are zero.
The inset plots the correlation energy of a mixture with $P$ species divided
by that of $P=2$ species, which we term the relative correlation energy.
(b) Same as (a) but for the correlation energy per species, $\frac{E_{cor}}{P}$, at the infinite-particle-number limit. 
See the text for further discussion.  
The quantities shown are dimensionless.}
\label{F1}
\end{figure}

\begin{figure}[!]
\begin{center}
\hglue -1.6 truecm
\includegraphics[width=0.43\columnwidth,angle=-90]{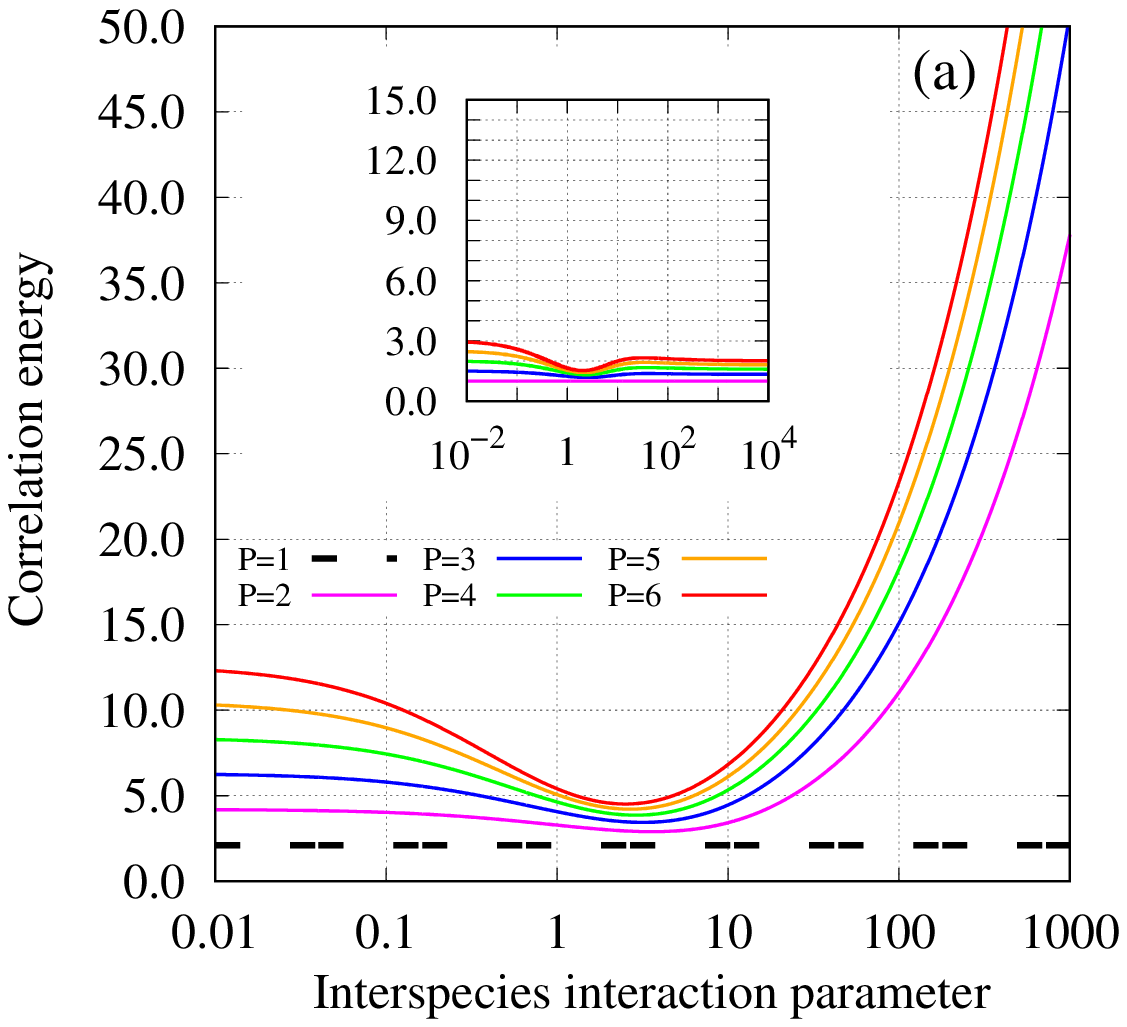}
\hglue -2.4 truecm
\includegraphics[width=0.43\columnwidth,angle=-90]{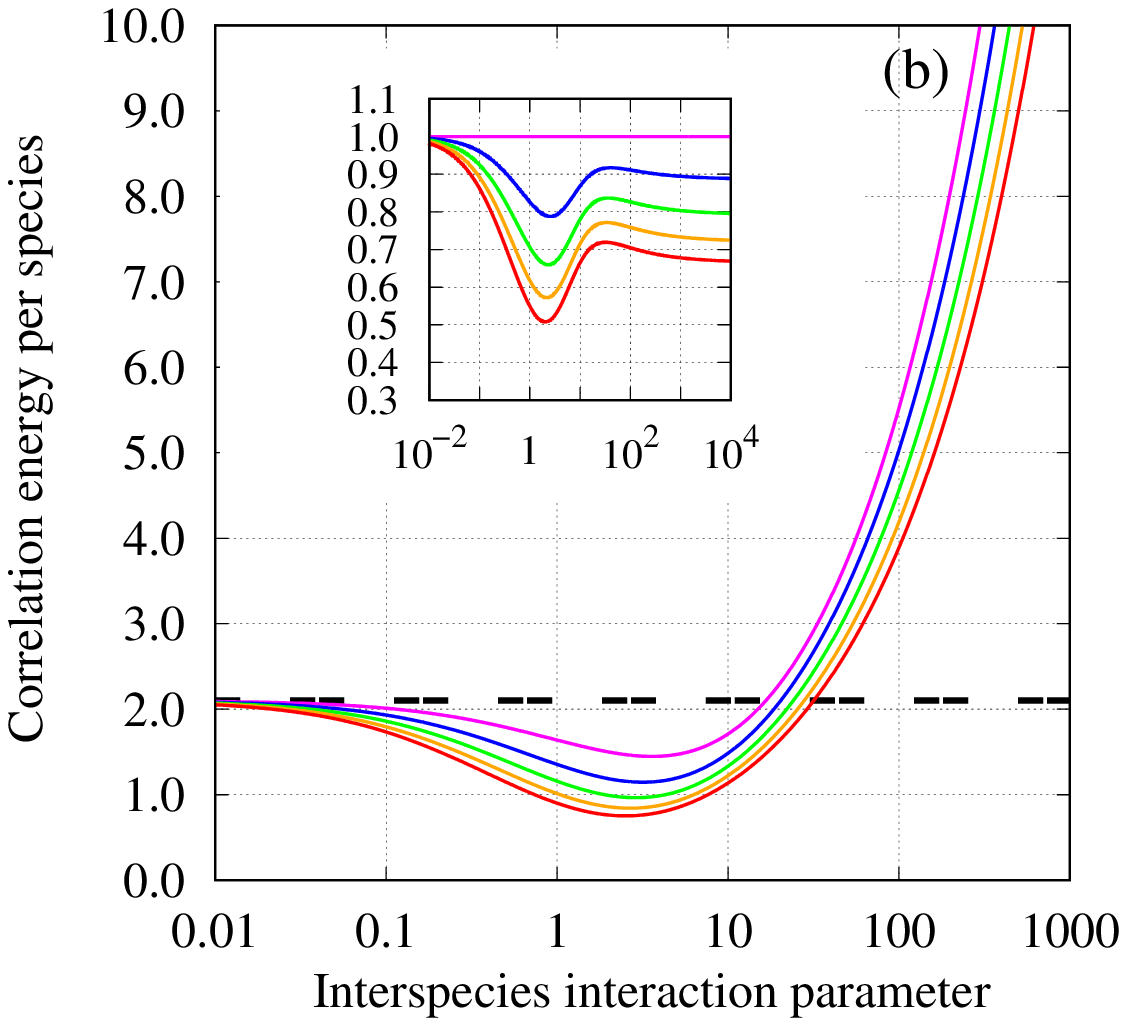}
\end{center}
\vglue 0.50 truecm
\caption{(a) Correlation energy, $E_{cor}$, and (b) correlation energy per particle, $\frac{E_{cor}}{P}$, at the infinite-particle-number limit
of $P$ species of bosons as a function of the interspecies interaction parameter $\Lambda_{12}$.
Same as Fig.~\ref{F1} but when the intraspecies interactions are $\Lambda_{1}=10$.
See the text for further discussion.  
The quantities shown are dimensionless.}
\label{F2}
\end{figure}

\begin{figure}[!]
\begin{center}
\hglue -1.6 truecm
\includegraphics[width=0.43\columnwidth,angle=-90]{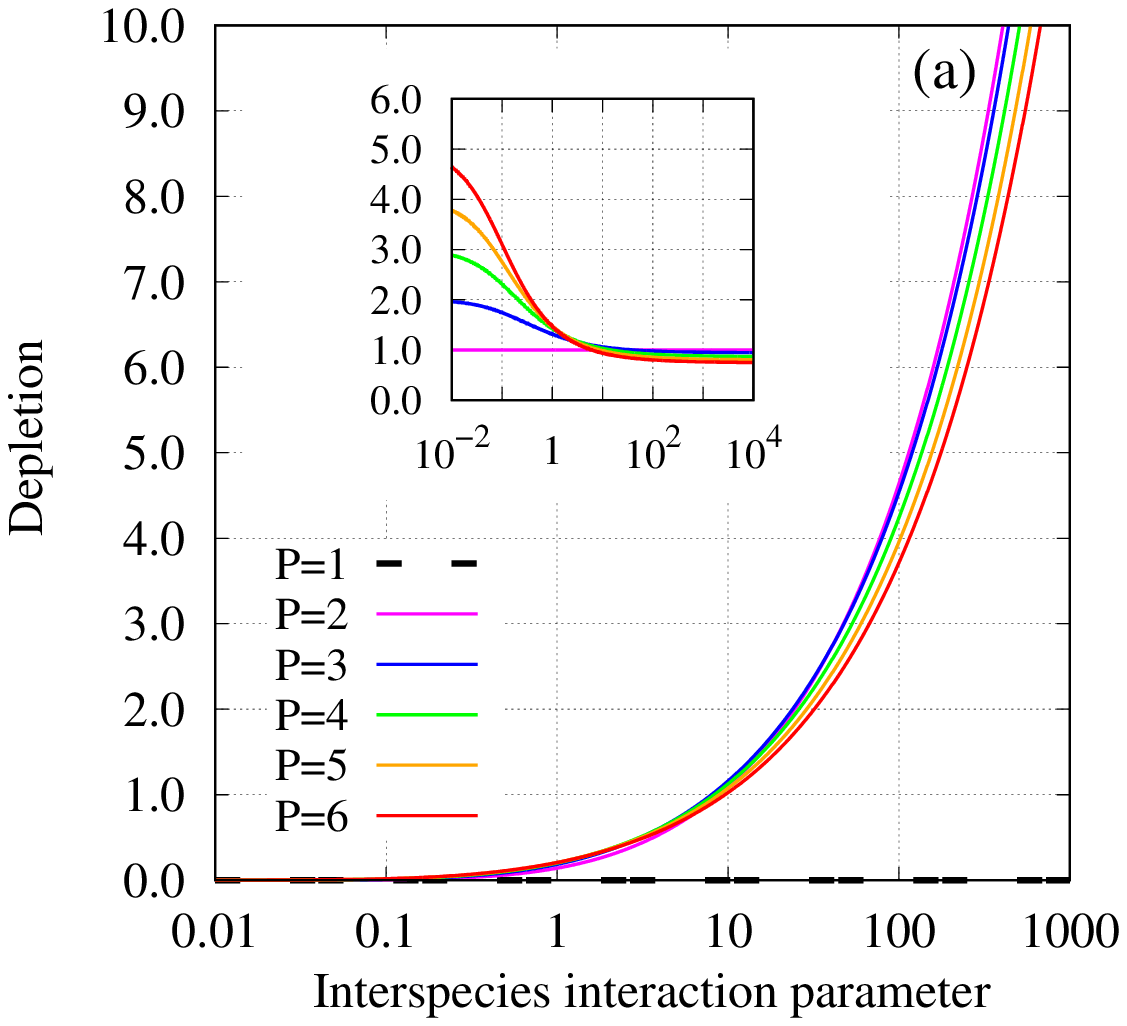}
\hglue -2.4 truecm
\includegraphics[width=0.43\columnwidth,angle=-90]{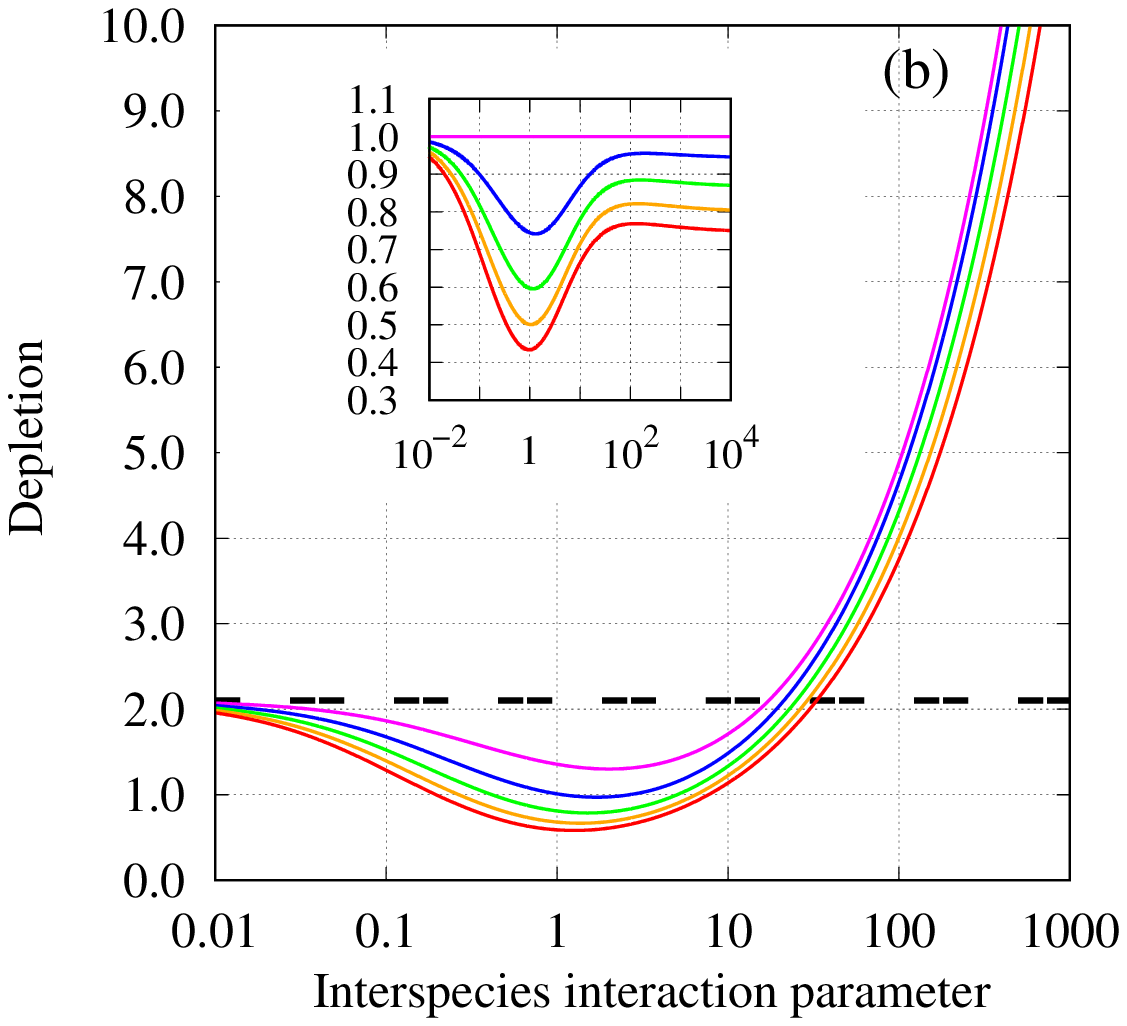}
\end{center}
\vglue 0.50 truecm
\caption{(a) Depletion, $d_1^{(1)}$, at the infinite-particle-number limit
of species $1$ bosons in the $P$-species mixture as a function of the interspecies interaction parameter $\Lambda_{12}$.
The inset depicts the depletion of a mixture with $P$ species divided
by that of $P=2$ species, which we term the relative depletion.
The intraspecies interactions are zero.
(b) Same as panel (a) but 
when the intraspecies interactions are $\Lambda_{1}=10$.
The depletion, $d_1^{(1)}$, and correlation energy per species, $\frac{E_{cor}}{P}$, behave similarly,
compare Fig.~\ref{F1}b to panel (a) and Fig.~\ref{F2}b to panel (b), respectively.
See the text for further discussion.  
The quantities shown are dimensionless.}
\label{F3}
\end{figure}

\begin{figure}[!]
\begin{center}
\hglue -1.6 truecm
\includegraphics[width=0.43\columnwidth,angle=-90]{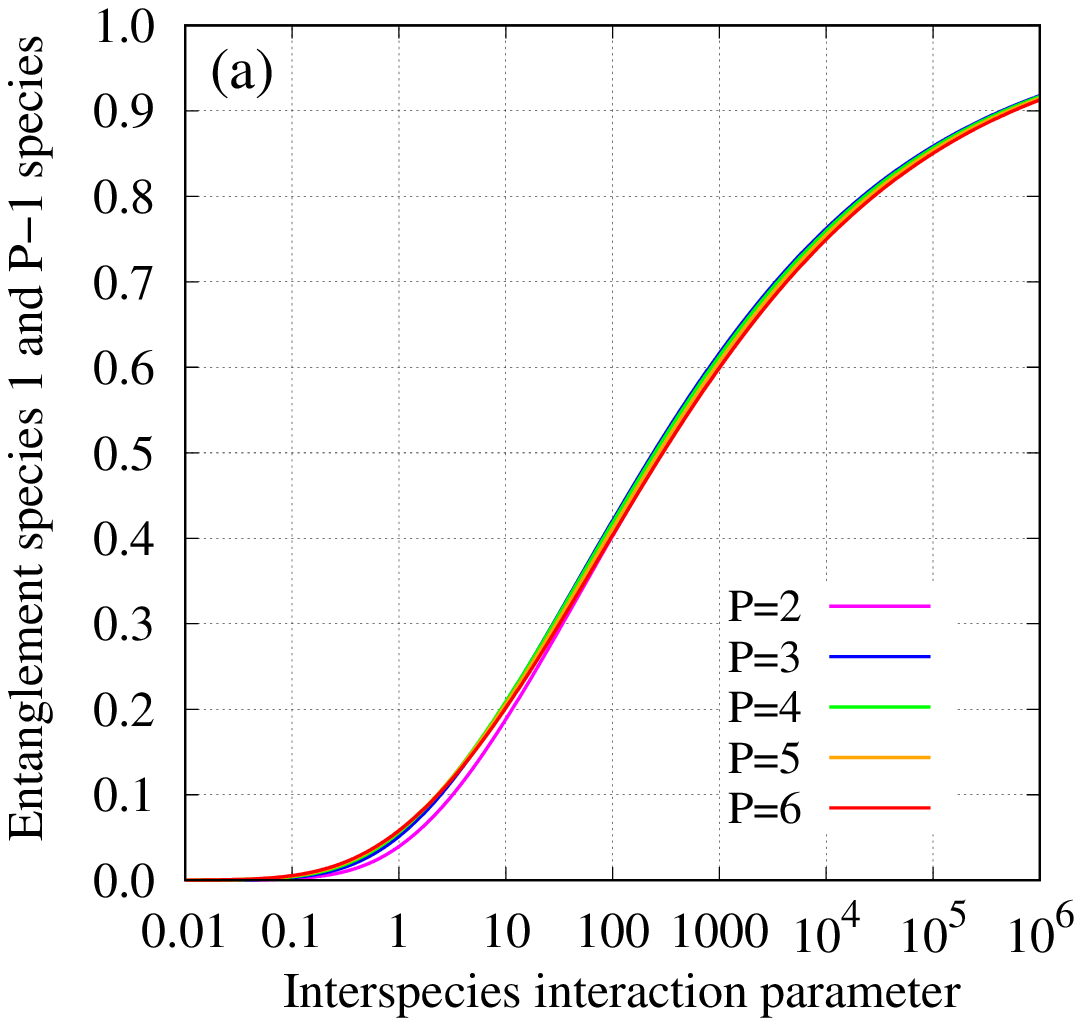}
\hglue -2.4 truecm
\includegraphics[width=0.43\columnwidth,angle=-90]{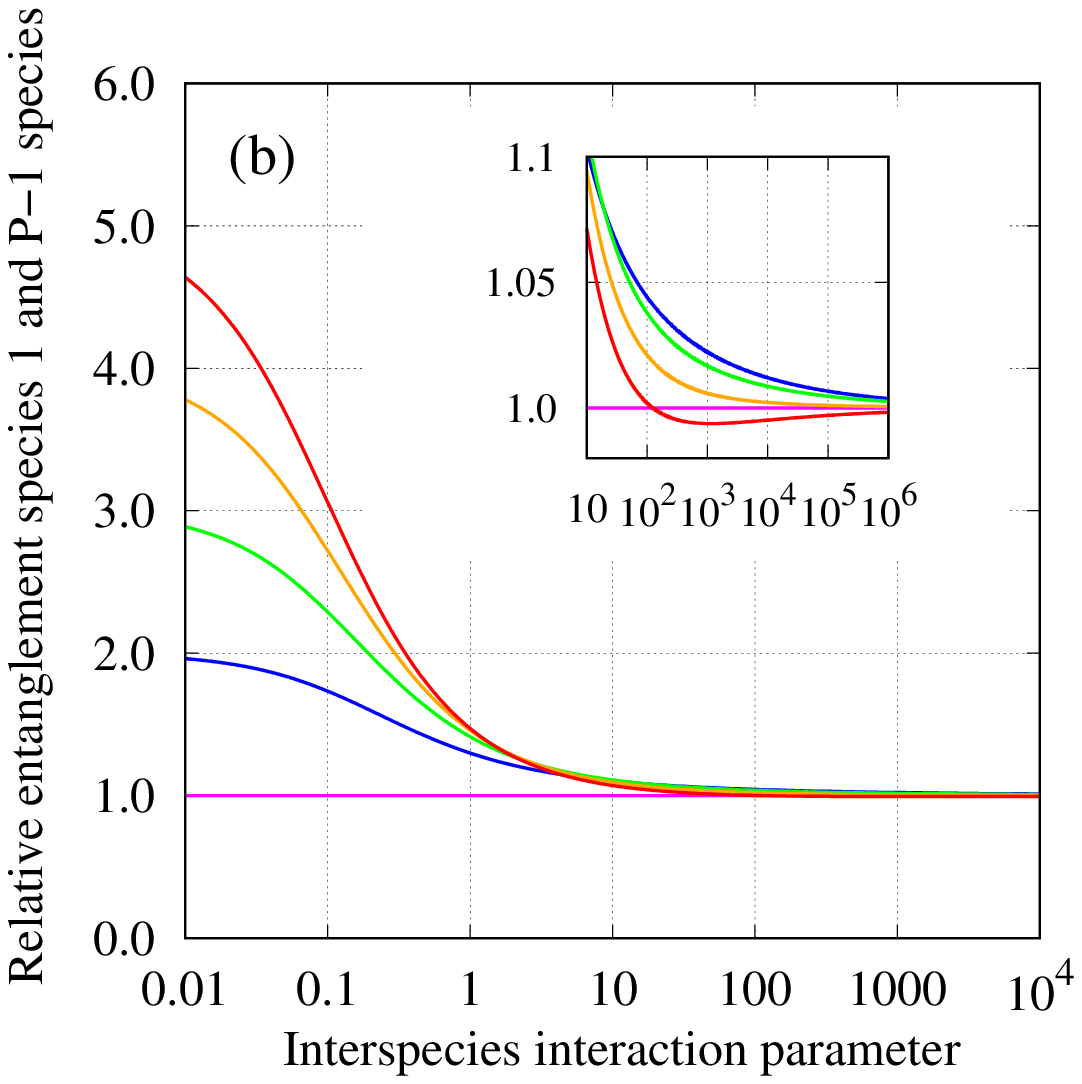}
\end{center}
\vglue 0.50 truecm
\caption{(a) 
Depicted is the Schmidt-decomposition parameter, $\rho^2$, defining the
von Neumann entanglement entropy  
between one species of bosons and the other $P-1$ species
at the infinite-particle-number limit as a function of the interspecies interaction parameter $\Lambda_{12}$.
(b) $\rho^2$
in a mixture with $P$ species divided
by that with $P=2$ species,
which for brevity we term the relative Schmidt-decomposition parameter (relative entanglement).
The inset zooms in on the results.
The Schmidt-decomposition parameter, $\rho^2$, follows the trend of the uncertainty product,
${\Delta^2}_{\hat X_{1,CM}} {\Delta^2}_{\hat P_{X_{1,CM}}}$,
compare Figs.~\ref{F4} and \ref{F5}.
The respective functional dependence between the two quantities is given in Eq.~(\ref{WSDrho_1}).
See the text for further discussion.  
The quantities shown are dimensionless.}
\label{F4}
\end{figure}

\begin{figure}[!]
\begin{center}
\hglue -1.6 truecm
\includegraphics[width=0.43\columnwidth,angle=-90]{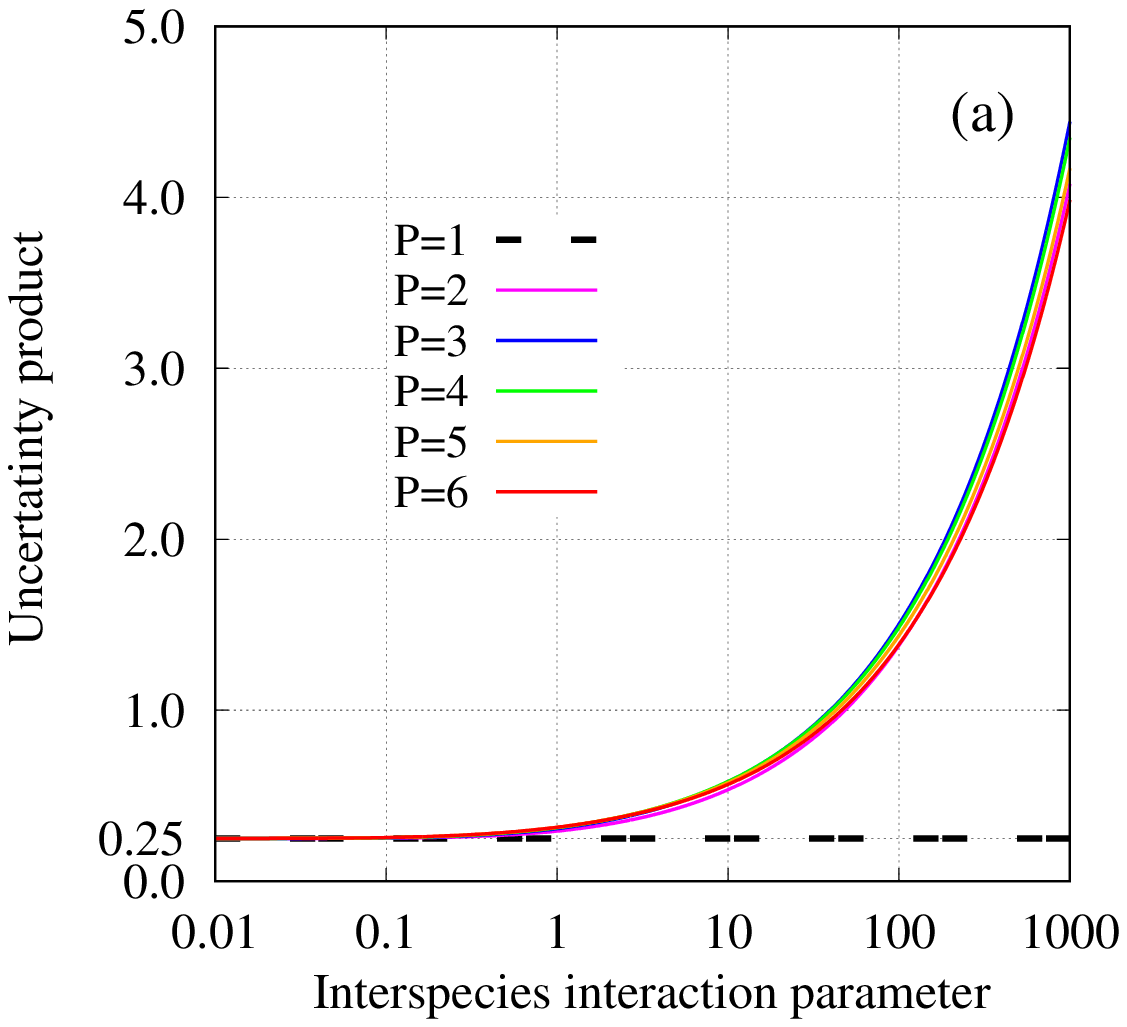}
\hglue -2.4 truecm
\includegraphics[width=0.43\columnwidth,angle=-90]{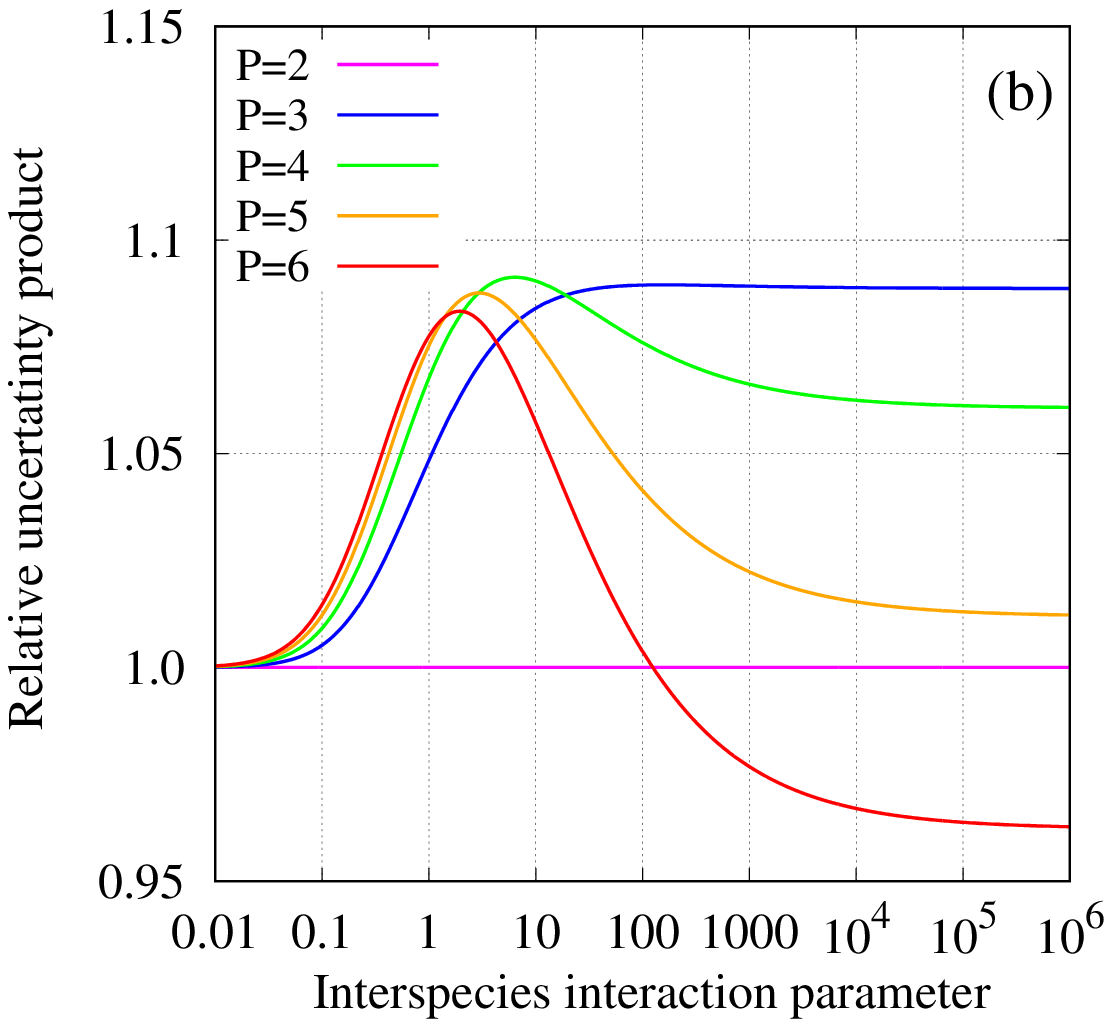}
\end{center}
\vglue 0.50 truecm
\caption{(a) Uncertainty product, ${\Delta^2}_{\hat X_{1,CM}} {\Delta^2}_{\hat P_{X_{1,CM}}}$, at the infinite-particle-number limit
of species $1$ bosons in the $P$-species mixture as a function of the interspecies interaction parameter $\Lambda_{12}$.
(b) Shown is the uncertainty product in a mixture with $P$ species divided
by that with $P=2$ species,
which we term the relative uncertainty product.
The uncertainty product, ${\Delta^2}_{\hat X_{1,CM}} {\Delta^2}_{\hat P_{X_{1,CM}}}$,
follows the trend of the Schmidt-decomposition parameter, $\rho^2$,
compare Figs.~\ref{F5} and \ref{F4}.
The respective functional dependence between the two quantities is given in Eq.~(\ref{rhoWSD}).
See the text for further discussion.  
The quantities shown are dimensionless.}
\label{F5}
\end{figure}

The results are collected in Figs.~\ref{F1}-\ref{F5} and analyzed hereafter.
In this section, 
we only discuss quantities at the infinite-particle-number limit of $P$-species mixtures.
Hence, the write-up of the symbol
$\lim_{N \to \infty \atop J=1,\ldots,P}$ can be suppressed for brevity in what follows,
with no source of confusion.

Fig.~\ref{F1}a depicts the correlation energy $E_{cor}$
as a function of the interspecies interactions $\Lambda_{12}$
for mixtures with $P=2$, $3$, $4$, $5$, and $6$ species.
For reference, the results of the single-species system, $P=1$, are given as well.
The intraspecies interactions vanish, $\Lambda_1=0$.
Two trends may be anticipated and indeed are immediately observed,
that the correlation energy increases with $\Lambda_{12}$ and 
with the number of species in the mixture $P$.
For the purpose of comparative analysis,
the inset of Fig.~\ref{F1}a plots the correlation energy of a mixture with $P$ species 
divided by that of the smallest mixture with two species,
which we term the relative correlation energy.
Corresponding to the correlation energy,
it is seen that the relative correlation energy
decreases monotonously with the interspecies interactions
and increases with the number of species.

To understand the behavior exhibited by the correlation energy,
it is useful to analyze the expressions of
$E_{cor}$ [Eq.~(\ref{E_cor_P_INF})]
for small and large interspecies interactions.
It is also instrumental
to define and discuss the correlation energy per species, $\frac{E_{cor}}{P}$.
Thus, for $\Lambda_{12} \gg \omega^2$ we find that
$E_{cor} \longrightarrow \frac{3}{\sqrt{2}} \sqrt{\Lambda_{12}} P\sqrt{P-1} \left(1-\sqrt{1-\frac{1}{P}}\right)$
and correspondingly
$\frac{E_{cor}}{P} \longrightarrow \frac{3}{\sqrt{2}} \sqrt{\Lambda_{12}} \sqrt{P-1} \left(1-\sqrt{1-\frac{1}{P}}\right)$
for the correlation energy per species.
On the other end,
for $\Lambda_{12} \ll \omega^2$,
we obtain that
$E_{cor} \longrightarrow \frac{3}{4}P(P-1) \frac{\Lambda_{12}^2}{\omega^3}$
and likewise
for the correlation energy per species $\frac{E_{cor}}{P}$.
Indeed, $E_{cor}$ increases with the number of species $P$ for weak and strong interspecies interactions
but $\frac{E_{cor}}{P}$ exhibits kind of a crossover,
as it increases for weak but deceases for strong $\Lambda_{12}$ with the number of species $P$ in the mixture.
These can be seen
in Fig.~\ref{F1} and its insets.

The picture changes when the intraspecies interactions are not zero.
Fig.~\ref{F2} depicts the results when $\Lambda_1=10$.
Now, the correlation energy is nonzero from the start,
and the effects of the interspecies interactions set in atop.
The general trend is that the correlation energy first decreases
with $\Lambda_{12}$ and than it increases unboundedly.
Analysis of $E_{cor}$ [Eq.~(\ref{E_cor_P_INF})]
for small and large interspecies interactions is deductive.
Hence, for $\Lambda_{12} \gg \omega^2, \Lambda_1$ we find that
the same limiting expressions as for $\Lambda_1=0$ before hold.
On the other end,
for $\Lambda_{12} \ll \omega^2, \Lambda_1$ we get that
$E_{cor} \longrightarrow \frac{3}{2}P\left[\left(\frac{\omega^2+\Lambda_1}{\sqrt{\omega^2+2\Lambda_1}}-\omega\right)
+(P-1)\left(\frac{\omega^2+3\Lambda_1}{\left(\omega^2+2\Lambda_1\right)^{\frac{3}{2}}}-\frac{1}{\omega}\right)\Lambda_{12}\right]$
and correspondingly for
$\frac{E_{cor}}{P}$.
Indeed, the pre-factor in front of $\Lambda_{12}$ is negative,
showcasing the initial decline of the correlation energy of the mixture
with the interspecies interactions.
Furthermore,
analysis of the correlation energy per species for both
weak and strong interspecies interactions
shows that $\frac{E_{cor}}{P}$ decreases with $P$.
Hence,
no crossover behavior like for $\Lambda_1=0$ occurs.
These features are all seen
in Fig.~\ref{F2} and its insets.

The correlation energy per species, $\frac{E_{cor}}{P}$,
is not only a more logical quantity to comparatively analyze than the correlation energy of the $P$-species mixture,
but, as we shall now see, is a predictor for the depletion of the species at the infinite-particle-number limit.
To remind, in the balanced mixture the depletion of each species is the same.
Indeed, comparing Fig.~\ref{F3} to the respective panels in Figs.~\ref{F1} and \ref{F2}
depicts a resemblance between the correlations at the infinite-particle-number limit 
exhibited by these two properties.
Again, analysis of the depletion
for $\Lambda_1=0$ and $\Lambda_1=10$
in the limits of weak and strong interspecies interactions
is useful here.
Thus we find,
for both $\Lambda_{12} \gg \omega^2$ and
$\Lambda_{12} \gg \omega^2, \Lambda_1$,
that $d_1^{(1)} \longrightarrow \frac{3}{2\sqrt{2}} \frac{\sqrt{P-1}}{P} \frac{\sqrt{\Lambda_{12}}}{\omega}$.
At the other end,
for $\Lambda_{12} \ll \omega^2$ one obtains
$d_1^{(1)} \longrightarrow \frac{3}{4}(P-1)\frac{\Lambda_{12}^2}{\omega^4}$
while for $\Lambda_{12} \ll \omega^2, \Lambda_1$ one gets
$d_1^{(1)} \longrightarrow
\frac{3}{4}\left[\frac{\left(\omega-\sqrt{\omega^2+2\Lambda_1}\right)^2}{\omega\sqrt{\omega^2+2\Lambda_1}} -
(P-1) \frac{4\Lambda_1^2}{\left[\omega^2\left(\omega^2+2\Lambda_1\right)\right]^{\frac{3}{2}}}\Lambda_{12}\right]$.
Indeed, for $\Lambda_1=0$
the depletion increases
and for $\Lambda_1=10$
it initially decreases with $\Lambda_{12}$.
In particular,
the depletion in the first case shows a crossover behavior,
since it increases for weak yet deceases for strong $\Lambda_{12}$ with the number of species $P$ in the mixture,
whereas in the second case no crossover behavior occurs. 
Thus, the properties of the depletion seen in Fig.~\ref{F3} are readily explained in these limits.

We now shift to the entanglement between one species and the remaining $P-1$ species
and its manifestation in an observable, the position--momentum uncertainty product.
To recall,
these properties are governed by the center-of-masses Hamiltonian (\ref{HAM_MIX_P_XYZWU_CMs}) and,
thus, do not depend on the intraspecies interactions $\Lambda_1$.
Figs.~\ref{F4} and \ref{F5}, respectively, present the results.
Since the Schmidt-decomposition parameter $\rho^2$ and 
the uncertainty product ${\Delta^2}_{\hat X_{1,CM}} {\Delta^2}_{\hat P_{X_{1,CM}}}$
are interconnected [Eqs.~(\ref{rhoWSD}) and (\ref{WSDrho_1})],
they follow the properties of each other.
It is useful to analyze the limiting cases of weak and strong interspecies interactions $\Lambda_{12}$.
Thus, we find for $\Lambda_{12} \gg \omega^2$ that
$\rho^2 \longrightarrow 1 - \left(\frac{1}{2\sqrt{2}}\frac{P-1}{P^{\frac{3}{2}}}\frac{\sqrt{\Lambda_{12}}}{\omega}\right)^{-\frac{1}{2}}$
and
${\Delta^2}_{\hat X_{1,CM}} {\Delta^2}_{\hat P_{X_{1,CM}}}
\longrightarrow \frac{1}{2\sqrt{2}}\frac{P-1}{P^{\frac{3}{2}}}\frac{\sqrt{\Lambda_{12}}}{\omega}$.
On the other end,
one has
$\rho^2 \longrightarrow \frac{1}{4}(P-1)\frac{\Lambda_{12}^2}{\omega^4}$
and
${\Delta^2}_{\hat X_{1,CM}} {\Delta^2}_{\hat P_{X_{1,CM}}}
\longrightarrow \left[1+(P-1)\frac{\Lambda_{12}^2}{\omega^4}\right]\frac{1}{4}$.
We can explicitly show now that a mixture with $P=3$ species exhibits,
relative to the mixtures with different numbers of species,
the maximal von Neumann entanglement entropy and uncertainty product
for strong interspecies interactions,
see Figs.~\ref{F4} and \ref{F5} and their insets.
The latter reflects the opposite effects
of (i) increasing the number of `neighbors' one species interacts with
and (ii) decreasing the relative mass of one species when the number
of `neighbors' is increased. 

In conclusion,
correlations at the infinite-particle-number limit
of a $P$-species mixture
show rich and appealing properties in comparison with single-species bosons,
when looking beyond the $100\%$ degree of condensation of each of its species.

\section{Summary and Outlook}\label{SEC_SUM_OUT}

We have presented in this work a solvable many-body model for a trapped multiple-species
mixture of Bose-Einstein condensates,
and, as an application, used it to construct a theory and analyze the correlation
properties of $P$-species bosonic mixtures at the infinite-particle-number limit.

We first obtained closed-form expressions for various quantities
as a function of the interactions and numbers of bosons.
These expressions will be used in future work to investigate properties
of multiple-species bosonic mixtures with a finite number of particles.
The infinite-particle-number limit is then performed explicitly.

As a starter,
the energy per particle, density per particle, and condensate fraction
are proven to boil down in the $P$-species mixture to their mean-field counterparts.
We then employ four properties to characterize correlations and
investigate how they change as a function of interactions
and, in particular, with the number of species in the mixture.
The properties investigated are
the correlation energy of the mixture, depletion of the species,
the von Neumann entanglement entropy between one species and the remaining $P-1$ species,
and the position--momentum uncertainty product of the species.
As a general trend,
correlations at the infinite-particle-number limit increase
with interspecies interactions,
although not monotonously if the intraspecies interactions are not zero.

The dependence of correlations at the infinite-particle-number limit
on the number of species $P$ is found to be intricate.
First, the behaviors of the correlation energy per species and of the depletion
follow a similar trend,
and, second,
the entanglement entropy and uncertainty product
are shown to be interconnected.
Chiefly, the correlation energy per species and the depletion
exhibit a crossover from weak to strong interspecies interactions
with the number of species $P$ (for zero intraspecies interactions).
For weak interactions, the larger is the number of species
the more correlated is the mixture,
whereas for strong interactions it is the other way around.
Furthermore,
the von Neumann entanglement entropy (Schmidt-decomposition parameter) and uncertainty product
also show a crossover from weak to strong interspecies
interactions.
Again, for weak interactions, the larger is the number of species
the more correlated is the mixture,
but for strong interactions we find an optimal number
of species, here $P=3$,
which maximizes the correlations in the mixture.
The latter represents the competition and balance
between interspecies interactions
and the intraspecies center-of-mass `weight' as $P$ is increased.

As an outlook,
we touch upon a couple of the many interesting possibilities that lie ahead.
One would naturally begin with the investigation of multiple-species
mixtures with finite numbers of particles.
Going beyond the balanced mixture,
generalizing the present treatment
to the generic multiple-species mixture would be rewarding.
Then, the different masses, intraspecies and interspecies interactions,
and the numbers of particles of the different species
is expected to enrich even more the properties of multiple-species mixtures with finite numbers of particles,
as well as at the infinite-particle-number limit(s).
In the latter case, e.g., some of the species could remain finite
while the rest become infinite.
A general question in a multiple-species
mixture is how its connectivity,
i.e.,
which species in the mixture interact with each other,
dictates properties.
We have seen above clues for partial 
insensitivity of the mean-field solution to the connectivity in the mixture,
and the issue calls for further investigations.
On the other end,
at the many-body level of theory,
the question of how entanglement between one group of non-interacting species
is transfered to a second group of non-interacting species
linked by a `bridge' between them would be interesting to explore.

Finally, in the longer run,
the present results and above prospected developments are expected to find applications
in modeling
Bose-Einstein condensates in cavities
and benchmarking coupled-cluster theory for bosonic mixtures
beyond the `two-species' treatments in \cite{HIM_MIX_CC, HIM_BEC_CAVITY},
respectively.
This is a good place to conclude the present work.

\section*{Acknowledgements}

We thank Alexej I. Streltsov for motivating discussions.
This research was supported by the Israel Science Foundation 
(Grant No. 1516/19). 

\appendix

\section{Revisiting the two-species mixture and its properties at the infinite-particle-number limit}\label{SEC_MIX2}

In this appendix,
we revisit the two-species mixture at the infinite-particle-number limit,
and collect results as far as they are needed
for comparison with and exposition of the main text.
To this end,
we apply the elimination of the other species' center-of-mass scheme to re-derive more concisely literature results,
and to augment them with derivation of the correlation energy and the depletion.

The two-species balanced-mixture Hamiltonian is given by
\beqn\label{HAM_MIX_2}
& & \hat H(\x_1,\ldots,\x_{N},\y_1,\ldots,\y_{N}) =
\sum_{j=1}^{N} \left( -\frac{1}{2} \frac{\partial^2}{\partial \x_j^2} + \frac{1}{2} \omega^2 \x_j^2 \right) + 
 \sum_{j=1}^{N} \left( -\frac{1}{2} \frac{\partial^2}{\partial \y_j^2} + \frac{1}{2} \omega^2 \y_j^2 \right) + 
\nonumber \\
& & + \lambda_1 \sum_{1 \le j < k}^{N} (\x_j-\x_k)^2 + 
\lambda_1 \sum_{1 \le j < k}^{N} (\y_j-\y_k)^2 + 
\lambda_{12} \sum_{j=1}^{N} \sum_{k=1}^{N} (\x_j-\y_k)^2. \
\eeqn

Introducing the Jacoby coordinates of each species 
\beqn\label{MIX_COOR_2}
& & \X_s = \frac{1}{\sqrt{s(s+1)}} \sum_{j=1}^{s} (\x_{s+1}-\x_j), \ \ 1 \le s \le N-1, \quad \X_{N} = \frac{1}{\sqrt{N}} \sum_{j=1}^{N} \x_j,  \nonumber \\
& & \Y_s = \frac{1}{\sqrt{s(s+1)}} \sum_{j=1}^{s} (\y_{s+1}-\y_j), \ \ 1 \le s \le N-1, \quad \Y_{N} = \frac{1}{\sqrt{N}} \sum_{j=1}^{N} \y_j, \
\eeqn
the Hamiltonian (\ref{HAM_MIX_2}) is written as a sum of two,
$\hat H = \hat H_{rels} + \hat H_{CMs}$.
The Hamiltonian of the relative motions is
\beqn\label{HAM_MIX_2_XY_rels}
& & 
\hat H_{rels}(\X_1,\ldots,\X_{N-1},\Y_1,\ldots,\Y_{N-1}) =
\sum_{j=1}^{N-1} \left( -\frac{1}{2} \frac{\partial^2}{\partial \X_j^2} + \frac{1}{2} \Omega_1^2 \X_j^2 \right) +
\nonumber \\
& &
+ \sum_{j=1}^{N-1} \left( -\frac{1}{2} \frac{\partial^2}{\partial \Y_j^2} + \frac{1}{2} \Omega_1^2 \Y_j^2 \right), \
\eeqn
where
\beqn\label{INTRA_FREQ_2}
\Omega_1 = \sqrt{\omega^2+2\left[\left(1+\frac{1}{N-1}\right)\Lambda_1+\Lambda_{12}\right]}.
\eeqn
The relative-motion frequency of each species, $\Omega_1$,
is affected by the intraspecies interaction and, of course,
by the interaction with the other species.

The Hamiltonian of the center-of-masses is
\beqn\label{HAM_MIX_2_XY_CMs}
& &
\hat H_{CMs}(\X_{N},\Y_{N}) =
- \frac{1}{2} \left(\frac{\partial^2}{\partial \X_{N_1}^2} + \frac{\partial^2}{\partial \Y_{N_2}^2}\right) +
\frac{1}{2}
\begin{pmatrix}
\X_{N} &
\Y_{N} \cr
\end{pmatrix}
\underline{\underline{\bf O}}
\begin{pmatrix}
\X_{N} \cr
\Y_{N} \cr
\end{pmatrix}, \nonumber \\
& &
\underline{\underline{\bf O}} = \left(\omega^2 + 4\Lambda_{12}\right) \underline{\underline{\bf I}}
-2\Lambda_{12}\underline{\underline{\bf 1}}. \
\eeqn
Diagonalizing the frequencies' matrix
$\underline{\underline{\bf O}}$ one finds the two eigenvalues and eigenvectors
associated with the mixture's
center-of-mass degrees-of-freedom,
\beqn\label{CM_2_freq}
& & \Omega_{12} = \sqrt{\omega^2 + 4\Lambda_{12}}, \quad \omega, \
\eeqn
and
\beqn\label{CM_2_vecs}
\Q_1 = \frac{1}{\sqrt{2}}\left(-\X_N+\Y_N\right),
\qquad
\Q_2 = \frac{1}{\sqrt{2}}\left(\X_N+\Y_N\right), \
\eeqn
respectively.
The latter are the Jacoby coordinates built from the
two center-of-mass Jacoby coordinates of the two species, $\X_N$ and $\Y_N$.
Both vectors, the relative center-of-mass motion $\Q_1$ and
the center-of-mass motion $\Q_2$ of the mixture do not depend on
the strength of the interaction between the two species.
The frequencies $\Omega_{12}, \omega$
and
eigenvectors $\Q_1, \Q_2$
are fundamental quantities in the present appendix,
and their properties for three-species and multiple-species mixtures
are investigated and utilized in the main text.

The ground-state wavefunction is then a product of
the relative motions and center-of-masses parts and takes on the separable form
\beqn\label{WF_HIM_2}
& & \Psi(\X_1,\ldots,\Y_1,\ldots,\Q_1,\Q_2) = 
\left(\frac{\Omega_1}{\pi}\right)^{\frac{3(N_1-1)}{2}}
\left(\frac{\Omega_{12}}{\pi}\right)^{\frac{3}{4}}
\left(\frac{\omega}{\pi}\right)^{\frac{3}{4}} \times \nonumber \\
& & \times
e^{-\frac{1}{2} \left(\Omega_1 \sum_{k=1}^{N-1} \X_k^2 + \Omega_1 \sum_{k=1}^{N-1} \Y_k^2\right)}
e^{-\frac{1}{2} \left(\Omega_{12} \Q_1^2 + \omega \Q_2^2\right)}
\eeqn
along with the ground-state energy
\beqn\label{E_HIM_2}
& &
\!\!\!\!\!\!\!\!\!\!\!\!
E = \frac{3}{2} \left[2 (N-1) \Omega_1 + \Omega_{12} + \omega \right] = \nonumber \\
& &
\!\!\!\!\!\!\!\!\!\!\!\!
= \frac{3}{2} \left[2(N-1) \sqrt{\omega^2+2\left[\left(1+\frac{1}{N-1}\right)\Lambda_1+\Lambda_{12}\right]}
+ \sqrt{\omega^2 + 4\Lambda_{12}} + \omega \right]. \
\eeqn
Recall that the two frequencies must be positive for the mixture to be bound.
Thus,
\beqn\label{FREQ_2_BOUNDS}
\Lambda_{12} > - \frac{\omega^2}{4}, \qquad
\Lambda_1 > - \left(1-\frac{1}{N}\right)\left(\frac{\omega^2}{2} + \Lambda_{12}\right).
\
\eeqn
Hence,
the energy (\ref{E_HIM_2}) is bound from below but not from above.

To proceed,
we work in the representation of the wavefunction using explicitly the
Jacoby coordinates of each species,
\beqn\label{WF_HIM_2_JACS}
& & \Psi(\X_1,\ldots,\X_{N},\Y_1,\ldots,\Y_{N}) =
\left(\frac{\Omega_1}{\pi}\right)^{\frac{3(N-1)}{2}}
\left(\frac{\Omega_{12}}{\pi}\right)^{\frac{3}{4}}
\left(\frac{\omega}{\pi}\right)^{\frac{3}{4}} \times \nonumber \\
& & \times
e^{-\frac{1}{2} \Omega_1 \sum_{k=1}^{N-1} \X_k^2}
e^{-\frac{1}{2} \Omega_1 \sum_{k=1}^{N-1} \Y_k^2}
e^{-\frac{1}{2} a \X_{N}^2}
e^{-\frac{1}{2} a \Y_{N}^2} 
e^{-b \X_{N}\Y_{N}}, \
\eeqn
where
\beqn\label{WF_HIM_2_JACS_COEFF}
& &
a = \frac{1}{2}\left(\Omega_{12}+\omega\right) = \Omega_{12}+b, \qquad
b = \frac{1}{2}\left(\omega-\Omega_{12}\right). \
\eeqn
The coefficients are interrelated and satisfy
$a+b=\omega$.
Their scaling properties at the limit of an infinite number of particles are discussed below,
and their
generalizations
to more species have been presented in the main text.

Then, the all-particle density matrix expressed using the species' Jacoby coordinates is
\beqn\label{N1_N2_DENS_JAC}
& & \Psi(\X_1,\ldots,\X_{N},\Y_1,\ldots,\Y_{N})\Psi^\ast(\X'_1,\ldots,\X'_{N},\Y'_1,\ldots,\Y'_{N}) = \nonumber \\
& & =
\left(\frac{\Omega_1}{\pi}\right)^{3(N-1)}
\left(\frac{\Omega_{12}}{\pi}\right)^{\frac{3}{2}}
\left(\frac{\omega}{\pi}\right)^{\frac{3}{2}}
e^{-\frac{1}{2} \Omega_1 \sum_{k=1}^{N-1} \left(\X_k^2+{\X'}_k^2\right)}
e^{-\frac{1}{2} \Omega_1 \sum_{k=1}^{N-1} \left(\Y_k^2+{\Y'}_k^2\right)} \times \nonumber \\
& & \times
e^{-\frac{1}{2} a \left(\X_{N}^2+{\X'}_{N}^2\right)}
e^{-\frac{1}{2} a \left(\Y_{N}^2+{\Y'}_{N}^2\right)}
e^{-b\left(\X_{N}\Y_{N}+\X'_{N}\Y'_{N}\right)}. \
\eeqn
Upon substitution of $a$ and $b$, and translating from Jacoby coordinates to the laboratory frame, 
the all-particle density matrix in \cite{HIM_MIX_RDM},
when considering
a balanced mixture \cite{Atoms_2021}, 
is recovered.

The integration of (\ref{N1_N2_DENS_JAC}) to arrive at the intraspecies reduced density matrices of species $1$ bosons
is performed in a few steps.
The first is eliminating the relative coordinates of species $2$ $\{\Y'_k=\Y_k, k=1,\ldots,N-1\}$.
The second step is integrating over the center-of-mass coordinate of species $2$ $\Y'_{N}=\Y_{N}$, i.e.,
\beqn\label{INT_CM_BB_2}
& & \int d\Y_{N} e^{-a \Y_{N}^2}
e^{-b\left(\X_{N}+\X'_{N}\right)\Y_{N}} =
\left(\frac{\pi}{a}\right)^{\frac{3}{2}} e^{+\frac{1}{4}\frac{b^2}{a}\left(\X_{N}+\X'_{N}\right)^2}. \
\eeqn
With this, all dependencies on the coordinates of species $2$ have been integrated out.
Translating the result from the Jacoby coordinates to the laboratory frame we arrive at the working expression
\beqn\label{F_1_0_BB_2}
& & e^{-\frac{\alpha}{2} \sum_{j=1}^{N} \left(\x_j^2+{\x'}_j^2\right) - \beta \sum_{1\le j <k}^{N}
\left(\x_j \x_k + \x'_j \x'_k\right)} e^{-\frac{1}{4}C_{N,0}\left\{\sum_{j=1}^{N}\left(\x_j+\x'_j\right)\right\}^2}, \
\eeqn
in which
all the information on the interaction with species $2$ is contained
in the three coefficients,
\beqn\label{F_1_0_BB_2_Coeff}
& & \alpha = \Omega_1 + \frac{1}{N}\left(a-\Omega_1\right) =
\Omega_1\left\{1+\frac{1}{N}\left[\frac{1}{2}\left(\frac{\omega}{\Omega_1}+\frac{\Omega_{12}}{\Omega_1}\right)-1\right]\right\}, \nonumber \\
& &
\beta = \alpha-\Omega_1 = 
\Omega_1\frac{1}{N}\left[\frac{1}{2}\left(\frac{\omega}{\Omega_1}+\frac{\Omega_{12}}{\Omega_1}\right)-1\right], \nonumber \\
& & C_{N,0} = - \frac{1}{N}\frac{b^2}{a} =
- \frac{1}{N}\frac{1}{2} \frac{\left(\omega-\Omega_{12}\right)^2}{\left(\omega+\Omega_{12}\right)}, 
\eeqn
where
the following relation has been used
$\sum_{s=1}^{N-1} \X_s^2 = \left(1-\frac{1}{N}\right) \sum_{j=1}^{N} \x_j^2 
-\frac{2}{N} \sum_{1 \le j < k}^{N} \x_j\x_k$.
$C_{N,0}$, as should be expected, is precisely the expression obtained in \cite{HIM_MIX_RDM},
when taking the balanced mixture,
which now allows one for
intraspecies integration of (\ref{F_1_0_BB_2}) following \cite{HIM_Cohen}.

The reduced one-particle density matrix then reads
\beqn\label{rho_1_BB_2}
& & \rho^{(1)}_1(\x,\x') = N \left(\frac{\alpha+C_{1,0}}{\pi}\right)^{\frac{3}{2}}
e^{-\frac{\alpha}{2}\left(\x^2+{\x'}^2\right)}e^{-\frac{1}{4}C_{1,0}\left(\x+\x'\right)^2}, \nonumber \\
& & \rho^{(1)}_1(\x) = N \left(\frac{\alpha+C_{1,0}}{\pi}\right)^{\frac{3}{2}}
e^{-\left(\alpha+C_{1,0}\right)\x^2}, \
\eeqn
with
\beqn\label{rho_1_BB_2_coeff}
& &
\!\!\!\!\!\!\!\!\!\!\!\!\!\!\!\!
\alpha + C_{1,0} = \left(\alpha-\beta\right) \frac{\left(\alpha-\beta\right) + N\left(C_{N,0}+\beta\right)}
{\left(\alpha-\beta\right) + (N-1)\left(C_{N,0}+\beta\right)} = \Omega_1
\frac{1}{1+\frac{1}{N}\left[\frac{1}{2}\left(\frac{\Omega_1}{\omega}+\frac{\Omega_1}{\Omega_{12}} \right) - 1 \right]}, \
\eeqn
where
$C_{N,0} + \beta = C_{N,0} + \alpha - \Omega_1 = \frac{1}{N}\left(\frac{2\omega\Omega_{12}}{\omega+\Omega_{12}} - \Omega_1\right)
= - \frac{2\omega\Omega_{12}}{\omega+\Omega_{12}}
\frac{1}{N} \left[\frac{1}{2}\left(\frac{\Omega_1}{\omega}+\frac{\Omega_1}{\Omega_{12}}\right)-1\right]$
and
$C_{N,0} + \alpha = \Omega_1\left(1-\frac{1}{N}\right)+\frac{1}{N}\left(\frac{a^2-b^2}{a}\right) =
\Omega_1\left(1-\frac{1}{N}\right)+\frac{1}{N}\left(\frac{2\omega\Omega_{12}}{\omega+\Omega_{12}}\right)$
are used.

The essence of the mean-field solution goes as follows.
The detailed derivation is given in \cite{HIM_MIX_RDM}.
The ansatz for the wavefunction is the separable product state
\beq\label{MIX_WAV_GP_2}
 \Phi^{GP}(\x_1,\ldots,\x_{N},\y_1,\ldots,\y_{N}) = 
\prod_{j=1}^{N} \phi^{GP}_1(\x_j) \prod_{k=1}^{N} \phi^{GP}_2(\y_k).
\eeq
Sandwiching the many-body Hamiltonian and using the variational principle,
the coupled Gross-Pitaevskii equations are obtained,
and their solution is given by
\beqn\label{MIX_GP_OR_2}
& & \phi^{GP}_1(\x) = \left(\frac{\Omega_1^{GP}}{\pi}\right)^{\frac{3}{4}}
e^{-\frac{1}{2}\Omega_1^{GP}\x^2}, \quad
\Omega_1^{GP} = \sqrt{\omega^2 + 2\left(\Lambda_1+\Lambda_{12}\right)}, \nonumber \\
& & \phi^{GP}_2(\y) = \left(\frac{\Omega_1^{GP}}{\pi}\right)^{\frac{3}{4}}
e^{-\frac{1}{2}\Omega_1^{GP}\y^2}. \
\eeqn
The two orbitals are
the same in the balanced system.
There is no demixing in this model, also see \cite{HIM_MIX_RDM}.
We mention in this context that
a modification of the harmonic-interaction model to emulate demixing of two Bose-Einstein condensates
is introduced and used in \cite{HIM_JPCS_2022}.
The chemical potentials are $\mu_1=\mu_2=
\frac{3}{2}\left(\Omega_1^{GP}+\frac{\Lambda_1+\Lambda_{12}}{\Omega_1^{GP}}\right)$.
The Gross-Pitaevskii energy per particle can be written as
\beqn\label{MIX_GP_E_N_2}
& &
\varepsilon^{GP} = 
\frac{E^{GP}}{2N} =
\frac{3}{2}\Omega_1^{GP}=
\frac{3}{2}\sqrt{\omega^2 + 2\left(\Lambda_1+\Lambda_{12}\right)}.
\eeqn
That the infinite-particle-number limit implies $100\%$
Bose-Einstein condensation and boiling down
to the mean-field quantities is shown now.

Thus, one finds for the energy of the mixture at the infinite-particle-number limit
\beqn\label{INF_E_2}
\lim_{N \to \infty \atop J=1,2} \frac{E}{2N} = \varepsilon^{GP},
\eeqn
and for the reduced one-particle density matrix of species $1$
\beqn\label{INF_1_RDMs_2}
\!\!\!\!\!\!\!\!
\!\!\!\!
\lim_{N \to \infty \atop J=1,2}
\frac{\rho^{(1)}_1(\x,\x')}{N} =
\phi^{GP}_1(\x) \left\{\phi^{GP}_1(\x')\right\}^\ast.
\eeqn
Clearly from (\ref{INF_1_RDMs_2}),
also the density per particle, which is the diagonal of the reduced one-particle density matrix,
$\rho^{(1)}_1(\x) \equiv \rho^{(1)}_1(\x,\x'=\x)$, per particle,
boils down to the respective mean-field density.
Finally, since the mixture is balanced per consideration,
or, alternatively,
the assignment of the labeling $1$ and $2$ to the species is arbitrary,
relation (\ref{INF_1_RDMs_2}) holds for species $2$ in the mixture as well.

Next, the correlation energy reads
\beqn\label{E_cor_2}
& &
\!\!\!\!\!\!\!\!\!\!\!\!\!\!\!\!\!\!\!\!
E_{cor}=E^{GP}-E=
\frac{3}{2} \Bigg[2N\sqrt{\omega^2 + 2\left(\Lambda_1+\Lambda_{12}\right)} - \nonumber \\
& &
\!\!\!\!\!\!\!\!\!\!\!\!\!\!\!\!\!\!\!\!
- 2(N-1) \sqrt{\omega^2+2\left[\left(1+\frac{1}{N-1}\right)\Lambda_1+\Lambda_{12}\right]}
- \sqrt{\omega^2 + 4\Lambda_{12}} - \omega \Bigg]. \
\eeqn
Using
$\Omega_1 = \sqrt{\omega^2+2\left[\left(1+\frac{1}{N-1}\right)\Lambda_1+\Lambda_{12}\right]}
= \Omega_1^{GP}\sqrt{1+\frac{2\Lambda_1}{(N-1)\{\Omega_1^{GP}\}^2}}$
one obtains at the infinite-particle-number limit
\beqn\label{E_cor_2_INF}
& &
\!\!\!\!\!\!\!\!
\lim_{N \to \infty \atop J=1,2} E_{cor}=
\frac{3}{2} \Bigg[2\sqrt{\omega^2 + 2\left(\Lambda_1+\Lambda_{12}\right)} -
\frac{2\Lambda_1}{\sqrt{\omega^2 + 2\left(\Lambda_1+\Lambda_{12}\right)}} -
\sqrt{\omega^2 + 4\Lambda_{12}} - \omega \Bigg] = \nonumber \\
& &
= \frac{3}{2} \left[ 2 \Omega_1^{GP} - \frac{2\Lambda_1}{\Omega_1^{GP}} - \Omega_{123} - \omega \right]. \
\eeqn
Expression (\ref{E_cor_2_INF}) tells us that, even at the limit of an infinite number of particles when
the two species are 100\% condensed and the many-body and mean-field energies per particle
coincide, there are correlations in the mixture.

The reduced one-particle density matrix (\ref{rho_1_BB_2},\ref{rho_1_BB_2_coeff})  
is diagonalized using Meheler's formula \cite{Robinson_1977,Schilling_2013,Atoms_2021}.
From which,
the final result for the depletion of species $1$ reads
\beqn\label{1RDMs_DEPLT_1}
& & d^{(1)}_1 = N\left[1-\left(1-\rho^{(1)}_1\right)^3\right] =
N\rho^{(1)}_1 \left(3-3\rho^{(1)}_1+\left\{\rho^{(1)}_1\right\}^2\right), \nonumber \\
& &
\rho^{(1)}_1 = \frac{\mathcal{W}-1}{\mathcal{W}+1},\qquad \qquad
\mathcal{W} = \sqrt{\frac{\alpha}{\alpha+C_{1,0}}} = \\
& &
= \sqrt{\left\{1 + \frac{1}{N} \left[\frac{1}{2}\left(\frac{\omega}{\Omega_1} +
\frac{\Omega_{12}}{\Omega_1}\right) - 1 \right]\right\}\left\{1 + \frac{1}{N} \left[\frac{1}{2}\left(\frac{\Omega_1}{\omega} + \frac{\Omega_1}{\Omega_{12}}\right) - 1 \right]\right\}}. \nonumber \
\eeqn

At the infinite-particle-number limit we obtain the depletion
\beqn\label{1RDMs_DEPLT_1_INF}
& &
\lim_{N \to \infty \atop J=1,2} d^{(1)}_1 =
\frac{3}{4}\left[\frac{1}{2}\frac{\left(\omega-\Omega_1^{GP}\right)^2}{\omega\Omega_1^{GP}} +
\frac{1}{2}\frac{\left(\Omega_{12}-\Omega_1^{GP}\right)^2}{\Omega_{12}\Omega_1^{GP}}\right]. \
\eeqn
In the absence of interaction between the two species,
(\ref{1RDMs_DEPLT_1_INF}) reduces to the single-species infinite-particle-number depletion, see \cite{INF_LENZ_2017}.

The Schmidt decomposition of the wavefunction (\ref{WF_HIM_2_JACS}) goes as follows, also see \cite{Atoms_2021}.
Since the Jacoby relative coordinates' parts of the wavefunction are separable,
we only need to Schmidt decompose the parts with the coupled intraspecies center-of-mass Jacoby coordinates.
We dub these parts of the
wavefunction as
the mixture's normalized center-of-masses wavefunction.
It is given along with the final result of the decomposition by
\beqn\label{CM_JAC_BB_2_t}
& &
\Psi_{CMs}(\X_{N},\Y_{N}) =
\left(\frac{a^2-b^2}{\pi^2}\right)^{\frac{3}{4}}
e^{-\frac{1}{2}a\left(\X^2_{N}+\Y^2_{N}\right)}
e^{-b\X_{N}\Y_{N}} = \nonumber \\
& &
= \sum_{n_1,n_2,n_3=0}^\infty \left(1-\rho^2\right)^{\frac{3}{2}} \rho^{n_1+n_2+n_3} 
\Phi_{n_1,n_2,n_3}(\X_{N};s) \Phi_{n_1,n_2,n_3}(\Y_{N};s), \
\eeqn
where
\beqn\label{CM_JAC_BB_2_SQUEEZE_MEHLER_t2}
& &
\!\!\!\!\!\!\!\!\!\!\!\!\!\!\!\!\!\!\!\!
s = \sqrt{a^2-b^2} = \sqrt{\Omega_{12}\omega},
\qquad
\rho = \frac{\mathcal{W}_{SD}-1}{\mathcal{W}_{SD}+1},
\qquad
\mathcal{W}_{SD} = \sqrt{\frac{a-b}{a+b}} = \sqrt{\frac{\Omega_{12}}{\omega}}, \
\eeqn
which holds for attractive interspecies interaction ($b<0$).
For repulsive interspecies interaction ($b>0$),
just take, say, $\Y_N \to - \Y_N$
in (\ref{CM_JAC_BB_2_t})
and
$\rho \to - \rho$ in (\ref{CM_JAC_BB_2_SQUEEZE_MEHLER_t2}).

Manifestation of that in an observable is, for instance, the position--momentum uncertainty product, i.e.,
deviations from the mean-field separable-solution result
reflect
the entanglement between the two species.
Inverting (\ref{CM_2_vecs}) one computes
${\Delta^2}_{\hat \X_{N}} =
\frac{1}{4}\left(\frac{1}{\Omega_{12}}+\frac{1}{\omega}\right)\!\bo$
and
${\Delta^2}_{\hat \P_{\X_{N}}} =
\frac{1}{4}\left(\Omega_{12}+\omega\right)\!\bo$
and consequently \cite{HIM_MIX_CP}
\beqn\label{UP_BB_2}
{\Delta^2}_{\hat \X_{CM}} {\Delta^2}_{\hat \P_{\X_{CM}}} =
{\Delta^2}_{\hat \X_{N}} {\Delta^2}_{\hat \P_{\X_{N}}} =
\left[1 + \frac{1}{4}\frac{\left(\omega-\Omega_{12}\right)^2}{\omega\Omega_{12}}\right] \frac{1}{4}\bo.
\eeqn
Indeed,
the many-body result for the uncertainty product (\ref{UP_BB_2}) reflects 
the entanglement between the two species, see (\ref{CM_JAC_BB_2_SQUEEZE_MEHLER_t2}).
Both are a consequence of the coupling between the species center-of-masses, see (\ref{CM_JAC_BB_2_t}),
which survives the infinite-particle-number limit.
Inversely,
the coupling between the center-of-masses
does not exist at the Gross-Pitaevskii level,
there is no entanglement between the two species,
and, consequently,
$\left\{{\Delta^2}_{\hat \X_{CM}}{\Delta^2}_{\hat \P_{\X_{CM}}}\right\}^{GP} = \frac{1}{4}\bo$.
Further discussion is provided in Sec.~\ref{BBB_P_SEC} of the main text.

\section{Folding $P-1$ center-of-mass coordinates in the $P$-species mixture}\label{SEC_FOLD}

For $P=1$ species,
there are obviously no other species to fold down.
Consequently,
\beqn\label{C_1}
& &
C_N=0, \
\eeqn
which appears in the single-species treatment \cite{HIM_Cohen}.
For $P=2$ species one gets
\beqn\label{C_2}
& &
C_{N,0}=-\frac{1}{N}\frac{b^2}{a},
\eeqn
see \cite{HIM_MIX_RDM} and (\ref{F_1_0_BB_2_Coeff}) in the previous appendix.
For $P=3$ species we found
\beqn\label{C_3}
& &
C_{N,0,0}=-\frac{1}{N}\left[\frac{b^2}{a} + \frac{\left(b-\frac{b^2}{a}\right)^2}{a-\frac{b^2}{a}}\right] = -\frac{1}{N}\frac{2b^2}{a+b},
\eeqn
see (\ref{F_1_0_0_BBB_3_Coeff_1}).
For $P=4$ species, a concrete calculation gives
\beqn\label{C_4}
& &
\!\!\!\!\!\!\!\!\!\!\!\!
C_{N,0,0,0}=-\frac{1}{N}\left\{\frac{b^2}{a} + \frac{\left(b-\frac{b^2}{a}\right)^2}{a-\frac{b^2}{a}} +
\frac{\left[\left(b-\frac{b^2}{a}\right)-\frac{\left(b-\frac{b^2}{a}\right)^2}{a-\frac{b^2}{a}}\right]^2}
{\left(a-\frac{b^2}{a}\right)-\frac{\left(b-\frac{b^2}{a}\right)^2}{a-\frac{b^2}{a}}}
\right\}
= -\frac{1}{N}\frac{3b^2}{a+2b},
\eeqn
and so on.
We can prove the general expression (\ref{C_N_0_0_0_P}) using mathematical induction.

For this,
we write
\beqn\label{D_P}
& &
\!\!\!\!\!\!\!\!\!\!\!\!
C_{N,0,\ldots,0} = -\frac{1}{N} D_P(a,b), \qquad D_P(a,b) = \sum_{p=1}^P d_p(a,b), \qquad P \ge 1, \nonumber \\
& &
\!\!\!\!\!\!\!\!\!\!\!\!
d_1(a,b)=0, \quad d_2(a,b)=\frac{b^2}{a}, \quad d_{p+1}(a,b)=d_p\left(a-\frac{b^2}{a},b-\frac{b^2}{a}\right), \ p \ge 2. \
\eeqn
Following (\ref{C_1},\ref{C_2},\ref{C_3},\ref{C_4}) assume that
\beqn\label{IND_D_P}
D_P(a,b) = \frac{(P-1)b^2}{a+(P-2)b}.
\eeqn
Then, one must prove
using (\ref{D_P},\ref{IND_D_P}) that the succeeding relation holds
\beqn\label{IND_D_P+1}
& &
\!\!\!\!\!\!\!\!\!\!\!\!\!\!\!\!\!\!\!\!
D_{P+1}(a,b) = D_P(a,b) + d_{P+1}(a,b) = D_P(a,b) + d_P\left(a-\frac{b^2}{a},b-\frac{b^2}{a}\right) = \nonumber \\
& &
\!\!\!\!\!\!\!\!\!\!\!\!\!\!\!\!\!\!\!\!
= D_P(a,b) + D_P\left(a-\frac{b^2}{a},b-\frac{b^2}{a}\right) - D_{P-1}\left(a-\frac{b^2}{a},b-\frac{b^2}{a}\right) =
\frac{Pb^2}{a+(P-1)b}.
\eeqn
Indeed, after some straightforward and not too lengthy algebra, Eq.~(\ref{IND_D_P+1}) is readily proved for $P \ge 2$.
For $P=1$ we have a single species, not a mixture, yet the general expression (\ref{IND_D_P}) holds as well,
i.e., it reduces to (\ref{C_1}).

\end{document}